\documentclass[aps,nobibnotes,two column,amssymb,secnumarabic,superscriptaddress,reprint]{revtex4-1}
\usepackage{amsmath}
\usepackage{amssymb}
\usepackage{bbm}
\usepackage{xcolor}
\usepackage{graphicx}
\usepackage{dcolumn}
\usepackage{bm}
\usepackage{hyperref}

\begin{document}
\title{Unified Superradiant phase transitions}
\author{Jie Peng}
\affiliation{Hunan Key Laboratory for Micro-Nano Energy Materials and Devices and School of
Physics and Optoelectronics, Xiangtan University, Hunan 411105, China}
\affiliation{Department of Physical Chemistry, University of the Basque Country UPV/EHU, Apartado 644, E-48080 Bilbao, Spain}

\author{Enrique Rico}
\affiliation{Department of Physical Chemistry, University of the Basque Country UPV/EHU, Apartado 644, E-48080 Bilbao, Spain}
\affiliation{IKERBASQUE, Basque Foundation for Science,
Maria Diaz de Haro 3, E-48013 Bilbao, Spain}

\author{Jianxin Zhong}
\affiliation{Hunan Key Laboratory for Micro-Nano Energy Materials and Devices and School of
Physics and Optoelectronics, Xiangtan University, Hunan 411105, China}

\author{Enrique Solano}
\affiliation{Department of Physical Chemistry, University of the Basque Country UPV/EHU, Apartado 644, E-48080 Bilbao, Spain}
\affiliation{IKERBASQUE, Basque Foundation for Science,
Maria Diaz de Haro 3, E-48013 Bilbao, Spain}
\affiliation{Department of Physics, Shanghai University, Shanghai 200444, China}

\author{I\~{n}igo L. Egusquiza}
\affiliation{Department of Theoretical Physics and History of Science,
University of the Basque Country UPV/EHU, Apartado 644, E-48080 Bilbao, Spain}

\begin{abstract}
 We prove, by means of a unified treatment, that the superradiant phase transitions of Dicke and classical oscillator limits of simple light--matter models are indeed of the same type. We show that the mean-field approximation is exact in both cases, and compute the structure and location of the transitions in parameter space. We extend this study to a fuller range of models, paying special attention to symmetry considerations. We uncover general features of the phase structure in the space of parameters of these models. 
\end{abstract}
\maketitle
\emph{Introduction.--}
Since the groundbreaking result of Dicke, in which he found a \emph{superradiant} state for a localized radiating gas \cite{PhysRev.93.99}, the very concept of superradiance has been actively investigated. In a superradiant state, the average photon number is macroscopic, in that the ratio $\langle a^\dag a\rangle/N$ is finite in the limit $N\to\infty$, where $N$ is the number of intervening atoms.
The existence of superradiant phase transitions (SPT) was rigourously shown \cite{hepp,wang} in the Tavis--Cummings model \cite{PhysRev.170.379}, which describes coupling of $N$ spins to a single bosonic mode with dipolar coupling, but restricts it the Rotating Wave Approximation (RWA). Those early works also considered the extension to the multimode case, beyond the RWA and for a multitude of analogous models \cite{PhysRevA.8.1440,PhysRevA.8.2517,zhang,PhysRevA.95.053854,guo,li,li1,GJC18}.

One more recent development is the realization that a similar phase transition is also present in a different limit, which in the Rabi model case corresponds to the photon frequency tending to zero compared to qubit frequency \cite{Bakemeier,PhysRevA.87.013826,hwang1,hwang2,liu1,lv}. This limit is understood as a classical oscillator \cite{Bakemeier}.
They are realized at zero temperature \cite{larson}, thus falling prima facie under the purview of Quantum Phase Transitions (QPT)  \cite{sachdev_2011} (see however the discussion in \cite{larson} to this point). Following the theoretical developments, there has also been keen experimental interest in realizing at least some of these transitions \cite{Baumann:2010aa,PhysRevLett.118.073001}.

In this letter we put forward a unifying treatment of SPTs in a wide class of models  \cite{br,chen,jorge,ion,jin,pj0,guxin,jorge1,KMLSN19}, including both the thermodynamical  and the classical oscillator limit. In all these models there is a competition between the free hamiltonian for  photons and spins, which tends to the normal phase, and an interaction that, when increased, produces the superradiant phase. Furthermore, in all these models the mean field approximation  for the bosons  \cite{PhysRevA.8.2517} will be exact in the relevant limit. The physical phases are then studied by analysing the extrema of a Landau potential. The order parameter at hand is the macroscopic boson expectation number. The symmetry or lack thereof of the underlying Hamiltonian will be apparent in the Landau potential, and the presence of second and first order superradiant phase transitions will be directly related to it.

\emph{Unified  analysis.--}\label{s3} We  present a unified treatment of the thermodynamic limit of the Dicke model and the classical oscillator model of the quantum Rabi model and SPTs thereof. Thus, we first consider the family of models ($\hbar=1$)
\begin{equation}\label{eq:dicke0}
H=\omega a^\dagger a+\sum_{i=1}^{N} \frac{g}{\sqrt{N}} \sigma_{ix}(a+a^\dagger)+\sum_i \Delta \sigma_{iz}.
\end{equation}
As usual, $a$ and $a^\dag$ denote annihilation and creation operators for a bosonic mode (the ``photon''), while the Pauli matrices also carry a spin index.
The parameters of this family are $\omega$ (photon energy), $\Delta$ (spin energy), $g$ (coupling strength) and $N$ (number of spins), together possibly with inverse temperature $\beta=1/k_B T$.
 We will henceforward denote this model as the Dicke--Quantum Rabi (DQR) model.

 In the DQR model  in the thermodynamic limit the order parameter is  the ratio $\langle a^\dag a\rangle/N$, which, if  finite in the limit $N\to\infty$ for either the thermal or the ground state is the diagnostic of the superradiant phase.

 We now endeavour to identify regions of parameter space that can be understood as the infinite level set of a function $C$ on parameter space and for which the  ratio $\langle a^\dag a\rangle/C$ is an order parameter, as first hinted at in \cite{larson}. We thus rewrite the Hamiltonian in the form
\begin{equation}\label{eq:dicke}
H=\Delta\sum_{i=1}^{N}[\frac{\omega C}{N\Delta} \frac{a^\dagger a}{C}+ \frac{g\sqrt{C}}{\sqrt{N}\Delta} \sigma_{ix}\frac{(a+a^\dagger)}{\sqrt{C}}+ \sigma_{iz}].
\end{equation}
If  indeed a transition is to be found, all three terms in the expression above must appear, since otherwise there will be no competition that accounts for the change of phase. Furthermore, they are to be examined under the hypothesis that  $\langle a^\dag a\rangle/C$ be an order parameter. This amounts to requiring that  in the limit $C\to\infty$ the effective couplings $\Omega=\omega C/N\Delta$ and $\gamma=g\sqrt{C}/\Delta\sqrt{N}$ tend to a finite value when $C\to\infty$.

Thus the region of parameter space of interest is given by the functions $C=\Omega\left(N\Delta/\omega\right)\to\infty$, with $\Omega$ a finite nonzero quantity, while at the same time $g^2/\omega\Delta$ is kept finite. One way of reaching this region is to take $N\to\infty$, while keeping finite all other parameters; this corresponds to the usual thermodynamical limit. An alternative, the classical oscillator limit, is achieved by considering $\Delta/\omega\to\infty$, as well as $g/\omega=(\gamma/\sqrt{\Omega})\sqrt{\Delta/\omega}\to\infty$, while $\gamma$ and $\Omega$ are finite, together with $N$.
This analysis can be extended to the finite temperature case \cite{sl}.

Further to the thermodynamical and the classical oscillator limits both being included in the infinite level set of the combination of parameters $N\Delta/\omega$, they are also determined there by the mean field approximation. Namely in that the mean field partition function $\tilde{Z}$ and the full partition function $Z$ will coincide in that limit. In a more concrete way,  the mean field reduced free energy $\tilde{f}=-\ln \tilde{Z}/\beta N\Delta$ and the full reduced free energy $f=-\ln Z/\beta N\Delta$ are equal up to  $O(\omega/N\Delta)$ \cite{sl}.

Following this result, the analysis of phases in terms of the mean field approximation is exact in the relevant limit $\omega/N\Delta\to0$ with finite $g^2/\omega\Delta$. For the case at hand, namely for hamiltonian \eqref{eq:dicke0}, the mean field partition function is computed to be $\tilde{Z}=\sqrt{N\Delta/\pi\beta\omega^2} \int_{-\infty}^{+\infty}\mathrm{d}u\,\exp\left[-\beta N \Delta \phi(u)\right]\,,$
where $\phi(u)=u^2-\ln\left[2\cosh\left(\beta\Delta\sqrt{1+4\gamma^2u^2}\right)\right]/\beta\Delta$ plays the role of a Landau potential. This last statement is the consequence the integral being well approximated by $\exp\left[-\beta N\Delta\phi(u_{\mathrm{min}})\right]$ in the limit $\beta N\Delta\to\infty$, with $u_{\mathrm{min}}$ the location of the global minimum of $\phi(u)$. In other words, $f=-\ln\tilde{Z}/\beta N\Delta\approx \phi(u_{\mathrm{min}})$.  Notice that  $\beta N\Delta=\beta \omega(N\Delta/\omega)$, so the limit we were considering from the outset implies the validity of the further approximation of the integral by Laplace's method, as long as $\beta\omega\neq0$.

The superradiant or normal character of each relevant phase is determined by the order parameter $\langle a^\dag a\rangle/(N\Delta/\omega)$, which is well approximated by the mean field prediction $\omega\partial_\omega f=\omega\partial_\omega\phi(u_{\mathrm{min}})$. Since $\omega\partial_\omega\gamma=-\gamma/2$ and $\gamma\partial_\gamma\phi(u)=u\phi'(u)-2 u^2$, we conclude
\begin{equation}
  \label{eq:orderparameter}
  \frac{\langle a^\dag a\rangle}{N\Delta/\omega}\approx -\frac{\gamma}{2}\partial_\gamma\phi(u_{\mathrm{min}})=u_{\mathrm{min}}^2\,.
\end{equation}
Thus, the system is in a normal phase if the global minimum of the Landau potential $\phi(u)$ is zero, and in a superradiant phase if the origin is no longer the global minimum.
The critical line separating the two regions is $\tanh(\beta\Delta)-\frac{1}{2\gamma_c^2}=0$. For couplings smaller than the critical coupling $\gamma_c(\beta\Delta)$ determined by this equation the system is in its normal phase, and for $\gamma>\gamma_c$ in a superradiant one. Close to criticality $u^2_{\mathrm{min}}$ is proportional to $\gamma-\gamma_c$, and the transition is second order.
Notice that in this unified treatment we have recovered both the $N\to\infty$ Dicke SPT \cite{PhysRevA.8.1440} and the Quantum Rabi classical oscillator SPT \cite{hwang1}.

\emph{Symmetry.--}
\label{sec:symm-cons} The fact that Hamiltonian \eqref{eq:dicke0} presents a $\mathbbm{Z}_2$ symmetry is reflected in $\phi(u)$ being an even function. This, together with continuity and  $u^2$ being the dominant term of $\phi(u)$ for large $|u|$, fully determines that the phase transition is second order.
To better establish the connection between the $\mathbbm{Z}_2$ symmetry and the Landau potential being an even function,  we consider a single mode inhomogenous anisotropic extension of the DQR model,
\begin{eqnarray}\label{id0}
H&=&\omega a^\dagger a+\sum_{i=1}^{N}[\frac{g_i}{\sqrt{N}}(a^\dagger\sigma_i^-+a\sigma_i^+)+\frac{g_i\lambda_i}{\sqrt{N}}(a\sigma_i^-+a^\dagger\sigma_i^+)\nonumber\\
&+&\Delta_i\sigma_{iz}].
\end{eqnarray}
We term this model inhomogeneous because the dipole coupling constants $g_i$ are different for different qubits, and similarly the spin energies $\Delta_i$ and the anisotropy parameters $\lambda_i$. These are anisotropies in that rotating and counterrotating terms are differently presented.
In the general case ($\lambda_i\neq0$ at least for one value of $i$),
this model has a $\mathbbm{Z}_2$ symmetry with non trivial group element $\Pi=\exp[i\pi(a
^\dag a+\sum_i^N(1+\sigma_{iz})/2)]$. Since, for coherent states $|\alpha\rangle$, $\langle\alpha|\Pi H\Pi|\alpha\rangle= U\langle-\alpha|H|-\alpha\rangle U^\dag$, with $U=U^\dag=\prod_{i=1}^N\sigma_i^z$ unitary, the mean field Landau potential is even in $\alpha$.

Let $\Delta$ be a positive generalized mean of the set $\{\Delta_i\}$. Then the mean field approximation is asymptotically exact as $N\Delta/\omega\to\infty$, as before, and again the phase diagram is determined by the mean field Landau potential. As shown in  \cite{sl} there are indeed in this limit SPT, all of them second order and of mean field type, with $\mathbbm{Z}_2$ symmetry breaking. For finite $N\Delta/\omega\gg1$, the mean field approach is still approximately valid, and the $\mathbb{Z}_2$ symmetry should be preserved. So the ground state will take the approximate form of a ``Schr\"{o}dinger cat''-like state for finite $N$ \cite{sl}. Notice that this model includes as a special case the DQR Hamiltonian \eqref{eq:dicke0}.
A very different symmetry is present if $\lambda_i=0$ for all $i$, i.e. for the inhomogeneous Tavis--Cummings model. Namely a $U(1)$ symmetry generated by the conserved quantity $a^\dag a+\sum_i^N(1+\sigma_{iz})/2$. In this case the Landau potential depends on $\alpha$ only through $|\alpha|^2$, reflecting the underlying continuous symmetry, and the  spontaneous symmetry breaking of the second order SPT will be accompanied by  the presence of a Goldstone mode \cite{sl}.

To stress the role of symmetry, consider the explicit breaking of  the $\mathbbm{Z}_2$ by the addition of  a static bias $\epsilon_i\sigma_{ix}$ in the previous Hamiltonian. In this case there is no SPT, as expected, but rather a continuous increase of the order parameter with the couplings \cite{sl}.

 \emph{1st order SPT without symmetry.--}
  \begin{figure}[htbp]
\center
\resizebox{1\columnwidth}{!}{%
  \includegraphics{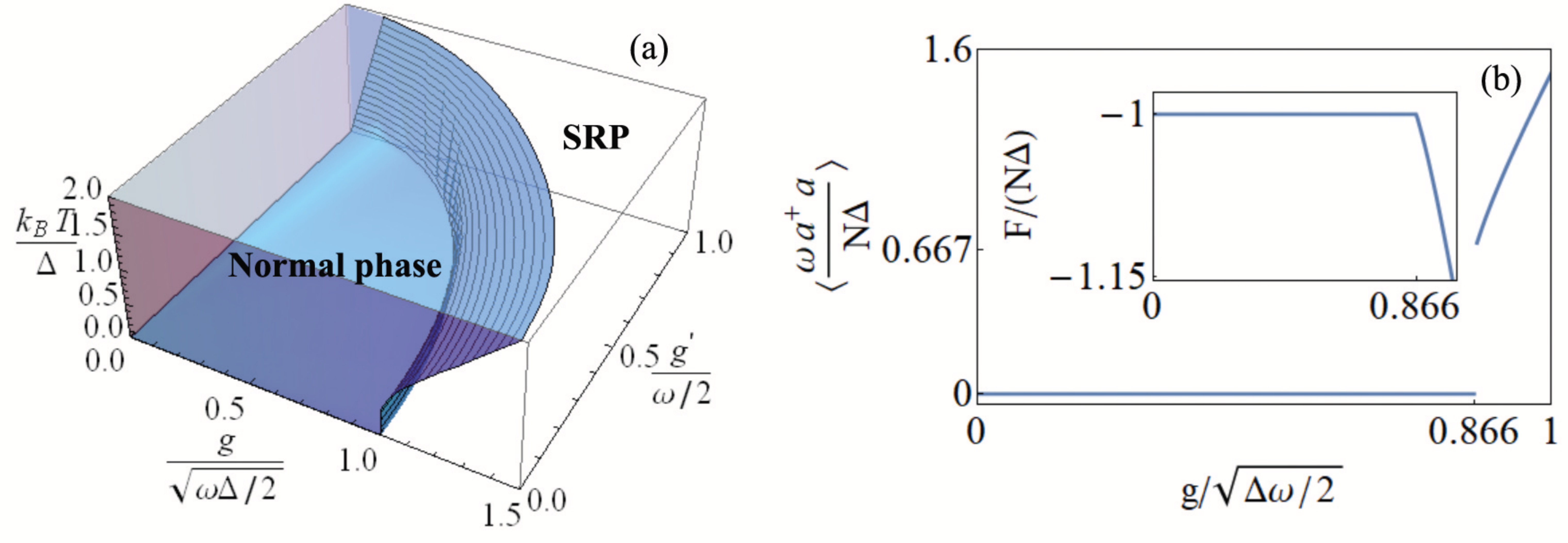}
}
\renewcommand\figurename{\textbf{Figure}}
\caption[2]{(a)Phase diagram of the Dicke model with one and two-photon terms in the thermodynamic limit. (b)The rescaled average photon number in the classical oscillator limit at finite temperature $\omega/k_B T\neq0$ with $g^\prime/\omega=0.25$, $\frac{g}{\sqrt{\Delta\omega/2}}=0.866$ and $\langle a^\dagger a\rangle_c/C=2/3$.   All the critical quantities are predicted analytically, and match  the numerical results.
\label{fig1}}
\end{figure}
A second order phase transition controlled by an order parameter requires a symmetry to be broken. Not so however for a first order transition to take place. As an example of this assertion in the context at hand, consider the QRM Hamiltonian  extended, homogeneously, with a two-photon coupling term \cite{ying,chen1},
\begin{equation}\small
\label{eq:Dicke12phtFirstHam}H=\omega a^\dagger a+\sum_{i=1}^{N}[ \sigma_{ix}[\frac{g}{\sqrt{N}} (a+a^\dagger)+ \frac{g^\prime}{N} (a^2+(a^\dagger)^2)]+\Delta \sigma_{iz}].
\end{equation}
The stability of the system requires $g'/\omega<1/2$ \cite{pj1,PhysRevA.92.033817}. In the generic case $g,g'\neq0$ this Hamiltonian has no symmetry. Following the by now well established routine, we obtain the reduced free energy as the global minimum of a potential
$\phi(u)=u^2 -\frac{1}{\beta\Delta}\ln\left[2\cosh\left(\beta\Delta\sqrt{1+\left[2\gamma u+2\gamma'u^2\right]^2}\right)\right]\,,$
where $\gamma=\frac{g}{\sqrt{\Delta\omega}}$ and $\gamma'=g^\prime/\omega$
\cite{sl}. The inexistence of the symmetry is reflected in the Landau potential not being even for the generic case. Nonetheless,  $u=0$ is always an extreme, and indeed the global minimum when $\gamma$ and $\gamma^\prime$ are close to $0$. On further analysis, there is a critical region in parameter space that separates normal and superradiant phases, determined by the existence of a nonzero solution of $\phi(u)=\phi(0)$ and $\phi'(u)=0$ simultaneously. When $\beta\Delta\to\infty$, this set of equations is algebraic and allows the determination of the explicit critical line $2\gamma^2+4\gamma'^2=1$, coinciding with the single qubit QPT case \cite{ying}. We present the phase diagram for the general case in Fig. \ref{fig1}(a). Given the lack of symmetry, the transition has to be first order, and this is indeed checked for $\beta\Delta\to\infty$, since just above the critical line  the rescaled photon number $u(\gamma_c,\gamma^\prime_c)_{\min}^2=\left(2\gamma^\prime_c/\gamma_c\right)^2\neq0$, as can be seen from Fig. \ref{fig1}(b).


 \emph{General structure of the phase diagram.--}
 \label{sec:inter-spins-first}
We now present a model with a discrete symmetry for which both continuous and discontinuous SPTs exist, and use it to illustrate general properties of the phase diagrams of the family of models under consideration.
In particular, we now study two spins which, on top of being dipole coupled to a photon mode, present XYZ spin-spin interactions \cite{pj2}, with $\mathbbm{Z}_2$ symmetry,
\begin{equation}
  \label{eq:2spinXYZ}\small
H=\omega a^\dagger a+\sum_{j=1,2}[g_j\sigma_{jx}(a+a^\dagger)+\Delta_j \sigma_{jz}]+\sum_{\alpha=x,y,x}J^{(\alpha)}\sigma_{1\alpha}\sigma_{2\alpha}.
\end{equation}
\begin{figure}[htbp]
\center
\resizebox{1\columnwidth}{!}{%
  \includegraphics{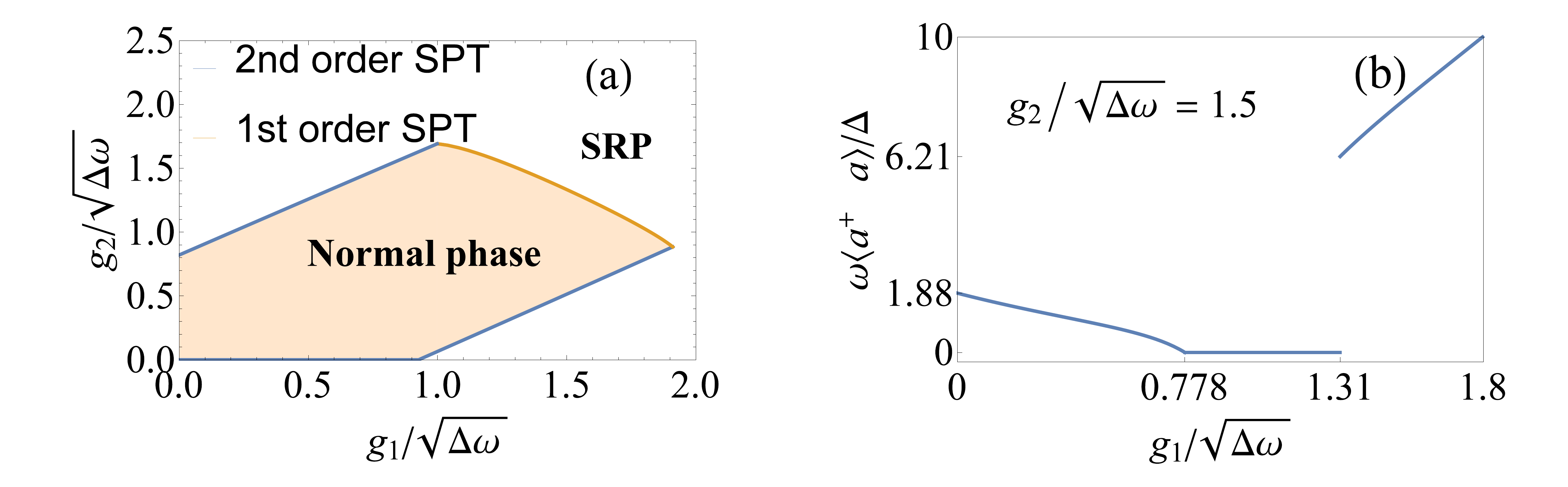}
}
\renewcommand\figurename{\textbf{Figure}}
\caption[2]{Nonidentical two-qubit Rabi model with XYZ spin-spin interactions. $\Delta_1/\Delta=3$, $\Delta_2/\Delta=2$, $J^x/\Delta=3$, $J^y/\Delta=2$, $J^z/\Delta=1$. (a)Phase diagram at any finite temperature $\omega/k_B T\neq0$. (b)Rescaled photon number for $\frac{g_2}{\sqrt{\Delta\omega}}=1.5$ at any finite temperature $\omega/k_B T\neq0$.
\label{fig2}}
\end{figure}
We study its phase diagram in  the classical oscillator limit $\Delta/\omega\to\infty$. Defining the reduced free energy $f$ from the partition function $Z=\exp\left(-\beta\Delta f\right)$, under the by now standard assumptions it will be determined as $\phi(u_{\mathrm{min}})$ at the global minimum of a Landau potential
  $\phi(u)=u^2+\lambda(u)$,
where $\lambda(u)$ is the smallest eigenvalue of the two spin operator
$h(u)=\sum_{j=1}^2\left[2\gamma_ju \sigma_{jx}+\delta_j\sigma_{jz}\right]+\sum_{\alpha=x,y,z}\epsilon_\alpha\sigma_{1\alpha}\sigma_{2\alpha}\,,$ where $\Delta$ is some positive generalised mean of $\{\Delta_j\}$, $\gamma_j=g_j/\sqrt{\Delta\omega}$, $\delta_j=\Delta_j/\Delta$, $\epsilon_\alpha=J^{(\alpha)}/\Delta$, and the order parameter $\omega\langle a^\dag a\rangle/\Delta$ is computed to be $u_{\mathrm{\min}}^2$.

The phase space of interest, thus, is seven dimensional, and we look at two dimensional sections controlled by the dipole couplings, organised in a vector $\vec{\gamma}$. In a generic section there will be both continuous and discontinuous phase transitions. The origin $\vec{\gamma}=0$ is always in normal phase, since the effective hamiltonian $h(u)$ is then independent of $u$. Furthermore, since the Landau function presents as $\phi(u)=u^2+f(u\vec{\gamma})$, with $f$ a function of several (here two) variables, we see that at critical points $u_*$ such that $\phi'(u_*)=0$ one has $\vec{\gamma}\cdot\nabla_\gamma\phi(u_*)=-2u_*^2$. Thus, for non-zero critical points, negative. Additionally, as $h(u)$ is a bounded operator, one sees that for large $\|\vec{\gamma}\|$ the minimum of $\phi$ can be approximately located at $|\gamma_1|+|\gamma_2|$, with the prediction that the system will be in its superradiant phase. It follows that there will be a region of normal phase around the origin, and that the radial component of the normal to its boundary, i.e. to the critical line, will not be zero.

For fixed $\vec{\delta}$, $\{\epsilon_\alpha\}$, one can identify whether the transition is first or second order in a given direction in the $\vec{\gamma}$ plane by several methods. First, the second derivative of $\phi(u)$ at zero can be obtained by a second order perturbative computation. If $\phi''(0)$ does not change sign as $\|\vec{\gamma}\|$ grows in a fixed direction, the transition in that direction will be necessarily discontinuous. Alternatively, since the origin in $u$ space cannot be the (unique) global minimum if there exists a point $u_s\neq0$ such that $\phi(0)=\phi(u_s)$, this equation is analysed. When this equation is inserted into the secular equation for $h(u)$ a cubic for the variable $u_s^2$ is obtained. The study of its discriminant provides us with criteria for the existence and  kind of solution, and thus for the location and character of the transition. In fact, the quantity $\phi''(0)$ will determine the value at the origin of the cubic, thus connecting both methods.

Notice that in any case there are symmetries in phase space on which we have relied in the discussion above. Thus, the effective spin hamiltonian $h(u)$ is isospectral with $h(-u)$ as a consequence of the $\mathbb{Z}_2$ symmetry of the initial Hamiltonian. This leads to the symmetry of phase space with respect to $\vec{\gamma}\to-\vec{\gamma}$. There are additional symmetries of phase space that can be identified in a similar manner. Thus, for instance, under the transformation $\vec{\delta}\to-\vec{\delta}$.

We illustrate this general analysis with the  portrayal of one $\vec{\gamma}$ plane section in Fig. \ref{fig2}.

We see in this manner that in the general class of spin models dipole coupled to a bosonic mode, if there is a symmetry, there will be a region of normal phase in the neighbourhood of the origin of the space of dipolar couplings, and that generally one flows radially in this space to a superradiant phase. The boundary will be composed generically of patches of first order and second order transition lines.

\emph{The multimode
  DQR model.-}
In the single mode analysis above we observe that the possiblity exists that by progressively increasing one dipole coupling the system can go from superradiant to normal to superradiant again. Thus, by using multiqubit systems we can have sequences of transitions, whose experimental realizations would be of great interest.

Even though the role of the bosonic mode and the spins are radically different, the fact above suggests that an extension to several bosonic modes (multi-mode in what follows) might provide us with rich phenomenology, even more amenable to experimental realization. Indeed  multimode models with dipole couplings can be analyzed with the techniques presented here \cite{sl}, and will have superradiant phases.

In particular, let us examine the following Hamiltonian: 
\begin{equation}\label{mode}
H=\sum_{\nu=1}^M\omega_\nu a_\nu^\dagger a_\nu+\sum_{i=1}^N\sum_{\nu=1}^M \frac{g_\nu}{\sqrt{N}} \sigma_{ix}(a_\nu+a^\dagger_\nu)+\sum_{i=1}^N \Delta_i \sigma_{iz}.
\end{equation}
In this case the dipole couplings are homogeneous for the qubits. Even if that were not the case, it is easy to observe that a $\mathbbm{Z}_2$ exists, with generator $\Pi=\exp\left\{i\pi\left[\sum_\nu a_\nu^\dag a_\nu+(1/2)\sum_i\left(1+\sigma_{iz}\right)\right]\right\}$. Furthermore, the system is in normal phase if the vector of dipolar couplings is zero. Following the steps of the previous analyses, we identify a set of order parameters $ \omega_\nu\langle a_\nu^\dag a_\nu\rangle/N\Delta$, with $\Delta$ some average of the spin frequencies $\Delta_i=\Delta \delta_i$. Similarly we arrive at a Landau function of $M$ variables $\{u_\nu\}_{\nu=1}^M$, whose global minimum determines the phase. In fact, these variable appear in the Landau function only through the combinations $\vec{u}^2$ and $\vec{\gamma}\cdot\vec{u}$, where we use vector notation for the corresponding $M$ dimensional objects, in the form $\phi(\vec{u})=\vec{u}^2-s\left[\left(\vec{\gamma}\cdot\vec{u}\right)^2\right]$ for some function $s$. The critical points of the Landau function, other than the origin, are therefore parallel to the vector $\vec{\gamma}$. Analogously to the analysis above, we see that for critical points other than the origin $\vec{\gamma}\cdot\nabla_\gamma\phi=-2\vec{u}^2$ is negative, whence the radial flow in coupling space is towards a superradiant phase. The coupling line is determined by $1/\vec{\gamma}_c^2=2\sum_i\tanh(\beta \Delta_i)/N|\delta_i|$. The transition is continuous, with the order parameters behaving radially with mean field exponent.
Since criticality is controlled by the length of the effective coupling vector $\vec{\gamma}$, one can achieve superradiance in a cooperative way. We term this phenomenon \emph{assisted} SPT.   

\emph{Conclusion and perspective.--}
We put forward the analytic study of SPT in a unified manner for both the thermodynamic and the classical oscillator limits of a wealth of boson-spin models with dipolar coupling. We show that in both cases the mean field approximation is exact, as was expected in the literature. Since we study the transitions in terms of a Landau potential, the analysis of symmetry comes to the fore, and we highlight its relevance in a general family of models. We demonstrate that general features of phase space in terms of dipolar couplings can be recovered from this approach, and portray anisotropies and first order/continuous boundaries. We expect that this general framework will be a good setting for new experimental proposals, given the wide range of phenomenology now available, including assisted SPT and the identification of first order transitions.

\emph{Acknowledgements.--} We are thankful to Ricardo Puebla for helpful discussions. This work was supported by the the program of China Scholarship Council (No. 201707230010), National Natural Science
Foundation of China (11704320, 11604240), Natural
Science Foundation of Hunan Province, China (
2018JJ3482), the National Basic Research
Program of China (2015CB921103), the Program for Changjiang Scholars and
Innovative Research Team in University (No. IRT13093), Spanish MINECO/FEDER FIS2015-69983-P,
Basque Government IT986-16, QMiCS (820505) and OpenSuperQ (820363) of the EU Flagship
on Quantum Technologies.

\newpage
\begin{widetext}

\subsection*{}
{\bf \large Supplementary Material: Superradiant phase transitions: exact unified presentation}
\renewcommand{\thesection}{S\arabic{section}}
\renewcommand{\thesubsection}{\Alph{subsection}}
\renewcommand{\thesubsubsection}{\alph\arabic{subsubsection}}
\renewcommand{\theequation}{S\arabic{equation}}
\renewcommand{\thefigure}{S\arabic{figure}}
\renewcommand{\thetable}{S\arabic{table}}
\setcounter{equation}{0}
\setcounter{figure}{0}

\section{Magnitude of parameters for SPT at finite temperature}
In the discussion of Eq. \eqref{eq:dicke} we have stressed the need for two parameters to be finite. Indeed, the finiteness of $\Omega$ and $\gamma$ has been predicated on our desire to identify a SPT at zero temperature. Nonetheless, even at finite temperature we would require their presence, as will be warranted by considering the partition function $Z=\mathrm{Tr}\left[\exp(-\beta H)\right]$ for the Hamiltonian in (\ref{eq:dicke}). The reduced free energy $f$, defined as $f=F/(N\Delta)= -(\ln Z)/(\beta N\Delta)$, should be finite for SPT to be  possible in all the (generalised) thermodynamic limits we examine. Denoting, as in the main text, the purportedly nonzero coefficients $\omega C/(N\Delta)=\Omega$ and $g\sqrt{C}/(\Delta\sqrt{N})=\gamma$, we find  that in a thermal state,
\begin{equation}
  \label{eq:delomf}
  \frac{\langle a^\dag a\rangle}{C}=\frac{\partial f}{\partial \Omega}\,,\qquad \frac{1}{N}\left\langle \sum_{i=1}^N \sigma_{ix}\frac{a+a^\dag}{\sqrt{C}}\right\rangle=\frac{\partial f}{\partial \gamma}\,.
\end{equation}
Since SPT takes place if the first quantity goes from zero in a region to non-zero elsewhere, the reduced free energy and its derivatives have to be finite, 
whence we require finiteness of $\Omega$ and $\gamma$ also in a thermal state.

\section{Validity of mean field approach on SPT}\label{sec:validity-mean-field}
Here we study the validity of the mean field approximation in
\begin{equation}\label{z11}
Z=\mathrm{Tr}\exp(-\beta H)=\int \mathrm{Tr}_{sp}\langle \alpha|e^{-\beta H(a^\dagger, a)}|\alpha\rangle\frac{d^2 \alpha}{\pi}\,,
\end{equation}
where $H(a^\dagger, a)=\omega a^\dagger a+\sum_{i=1}^{N}\left[ \frac{g}{\sqrt{N}} \sigma_{ix}(a+a^\dagger)+\Delta \sigma_{iz}\right]$.

Hepp and Lieb \cite{PhysRevA.8.25170} obtained the upper and lower bound
\begin{equation}\label{boundhepp1}
\tilde{Z}\leq Z\leq\exp\{\beta\omega\}\tilde{Z},
\end{equation}
where $\tilde{Z}=\int \mathrm{Tr}_{spin}\exp\{-\beta\left[ \omega|\alpha|^2+\sum_{i=1}^{N} \frac{g}{\sqrt{N}} \sigma_{ix}(\alpha+\alpha^*)+\sum_{i=1}^N\Delta \sigma_{iz}\right]\}\frac{d^2\alpha}{\pi}$ is the mean field approximation of $Z$. Here we proceed to prove the asymptotic validity of the mean field approximation when  ${N\Delta}/{\omega}$  tends to infinity, while maintaining ${g}/{\sqrt{\Delta\omega}}$ finite.

After some rearrangements and changing variables to  $\alpha=\sqrt{{N\Delta}/{\omega}}\alpha^\prime$, we obtain
  \begin{eqnarray}
 \tilde{Z}=\frac{N\Delta}{\omega}\int \mathrm{Tr}_{spin}\exp\{-\beta N\Delta h(\alpha^\prime,\alpha^{\prime*})\} \frac{d^2\alpha^\prime}{\pi},
 \end{eqnarray}
 where the effective spin hamiltonian  $h(\alpha^\prime,\alpha^{*\prime})$ is defined as $|\alpha^\prime|^2+\sum_{i=1}^{N} ({g}/{N\sqrt{\omega\Delta}}) \sigma_{ix}(\alpha^\prime+\alpha^{\prime*})+{\sum_{i=1}^N \sigma_{iz}}/{N}$.
 After tracing out the qubit part, $\tilde{Z}=({N\Delta}/{\omega})\int \exp\{-\beta N\Delta\phi_1(\alpha^\prime,\alpha^{*\prime})\} {d^2\alpha^\prime}/{\pi}$, with all coefficients of $\alpha^\prime$ and $\alpha^{*\prime}$ in $\phi_1(\alpha^\prime,\alpha^{*\prime})$ finite in the limit of interest.
 
 The bounds in  Eq. \eqref{boundhepp1} can be rewritten to 
 \begin{equation}
  \frac{-\ln \tilde{Z}}{\beta N\Delta}\geq \frac{-\ln Z}{\beta N\Delta }\geq -\frac{\omega}{N\Delta}-\frac{\ln \tilde{Z}}{\beta N\Delta } .
\end{equation}
Thus, if indeed $f=-{\ln \tilde{Z}}/{\beta N\Delta }$ is finite in the limit ${N\Delta}/{\omega}\to\infty$, the reduced free energy ${F}/{N\Delta}= {-\ln Z}/{\beta N\Delta}$  is asymptotically determined by $f$. For all finite values of $\beta\omega$, it is the case that $f$ is asymptotically given by the value of $\phi_1$ at its minimum, whence the result follows.

This analysis can be extended to multimode models with the form
\begin{equation}
H=\sum_{\nu}\left[ \omega_\nu a_{\nu}^\dag a_{\nu}+ a_{\nu}^\dag B+a_{\nu}^\dag B^\dag\right]+\sum_i\Delta\sigma_{iz},
\end{equation}
where B is a atomic operator, i.e. an operator acting on a finite dimensional Hilbert space. The bounds on their corresponding partition function are \cite{PhysRevA.8.25170}
\begin{equation}
\tilde{Z}\leq Z\leq \exp(\beta\sum_{\nu}\omega_{\nu})\tilde{Z}.
\end{equation}
We again observe that the mean field approximation is asymptotically exact as $\sum_\mu\omega_\mu/N\Delta\to0$.

Next we extend the study on bounds \cite{PhysRevA.8.25170} to the new case of interaction terms with  two photons,
\begin{equation}
H=\omega a^\dagger a+A+a^\dag B+a B^\dag+a^{\dag^2} D+a^2 D^\dag.
\end{equation}
$B$ and $D$ are again operators acting on a finite dimensional space $\mathfrak{H}$.

Following similar procedures to those presented in  \cite{PhysRevA.8.25170}, we define the cutoff coherent state as
\begin{equation}
|\alpha,n\rangle=P_n|\alpha\rangle,
\end{equation}
where $P_n$ is the projector onto the states with $n$ photons, so that $P_n\to I$ strongly. It is useful to define $K_n$ and identify some formulae in which it appears, as follows;
\begin{eqnarray}
&\langle \alpha,n|\beta,n\rangle=K_n(\alpha,\beta)=\exp[-\frac{1}{2}(|\alpha|^2+|\beta|^2)]\sum_{m=0}^n(\alpha^*\beta)^m/m!,\\
&\langle \alpha,n|a|\alpha,n\rangle=\alpha K_{n-1}(\alpha,\alpha),~\langle \alpha,n|a^\dag|\alpha,n\rangle=\alpha^* K_{n-1}(\alpha,\alpha),
&\langle \alpha,n|a^\dag a|\alpha,n\rangle=|\alpha|^2 K_{n-1}(\alpha,\alpha),\\
&\langle \alpha,n|a^2|\alpha,n\rangle=\alpha^2 K_{n-2}(\alpha,\alpha),~\langle \alpha,n|a^{\dag^2}|\alpha,n\rangle=\alpha^{*^2} K_{n-2}(\alpha,\alpha).
\end{eqnarray}
The standard convexity inequality $\langle \psi|e^X|\psi\rangle\geq\langle \psi|\psi\rangle\exp(\langle \psi|X|\psi\rangle/\langle \psi|\psi\rangle)$, or its more general Schwartz inequality form $P_n\exp(X)P_n\geq \exp(P_n XP_n)$ \cite{Davis:1957wj} provide us with the corresponding bound with cutoffs
\begin{eqnarray}
Z_n=\mathrm{Tr}\left[ P_n e^{-\beta H} P_n\right]\geq \mathrm{Tr}_\mathfrak{H}\int \frac{d^2\alpha}{\pi} \,K_n(\alpha,\alpha)
\times \exp\{-|\alpha|^2K_{n-1}(\alpha,\alpha)/K_{n}(\alpha,\alpha)+A\nonumber\\
+[B\alpha^*K_{n-1}(\alpha,\alpha)/K_{n}(\alpha,\alpha)+H.C.]
+[D\left(\alpha^*\right)^2K_{n-2}(\alpha,\alpha)/K_{n}(\alpha,\alpha)+H.C.]\}
\end{eqnarray}
Then we take the limit $n\to\infty$ in both sides to obtain $Z\geq\tilde{Z}$, where $\tilde{Z}$ is the partition function in mean field approximation.

On the other hand \cite{PhysRevA.8.25170,lieb},
\begin{equation}
Z_n(\epsilon)\leq \pi^{-1}\int \mathrm{d}^2 \alpha\, K_n(\alpha,\alpha)\mathrm{Tr}_\mathfrak{H}\exp\{-\beta H_n(\alpha,\epsilon)\},
\end{equation}
where
\begin{eqnarray}
H_n=P_n H P_n=\frac{1}{\pi}\int  \mathrm{d} \alpha H_n(\alpha)|\alpha,n\rangle\langle\alpha,n|,\nonumber\\
H_n(\alpha,\epsilon)=H_n(\alpha)e^{-\epsilon|\alpha|^2}.
\end{eqnarray}
Here
\begin{equation}
H_n(\alpha)=\omega(|\alpha|^2-1)+A+B\alpha^*+B^\dagger\alpha+D\alpha^{*^2}+D^\dag\alpha^{2}.
\end{equation} We then take the limit $n\to\infty$ and $\epsilon\to0$ to obtain $Z\leq e^{\beta\omega}\tilde{Z}$. Implicit in this analysis is the requirement that $\omega^2/4>\|D^\dag D\|$, as required for stability.

In conclusion, the bounds $\tilde{Z}\leq Z\leq e^{\beta\omega}\tilde{Z}$ do hold for  the (stable) two-photon interaction case, and the reduced free energy is given asymptotically by the mean field value.

\section{2nd order SPT with symmetry: anisotropic inhomogeneous DQR model}
Here we consider the anistropic inhomogeneous Dicke-quantum Rabi model,
\begin{eqnarray}\label{id}
H=\omega a^\dagger a+\sum_{i=1}^{N}\left[\frac{g_i}{\sqrt{N}}(a^\dagger\sigma_i^-+a\sigma_i^+)+\frac{g_i\lambda_i}{\sqrt{N}}(a\sigma_i^-+a^\dagger\sigma_i^+)
+\Delta_i\sigma_z\right].
\end{eqnarray}

As before, we consider the limit $N{\Delta}/{\omega}\to\infty$, in which case  the mean field approximation of the partition function $\tilde{Z}=\int({\mathrm{d}^2 \alpha}/{\pi})\, \mathrm{Tr}_{spin}\left[ e^{-\beta\langle\alpha|H(a^\dagger\to\alpha^*, a\to\alpha)|\alpha\rangle}\right]$ is asymptotically exact. Tracing out the qubit part, we obtain $\tilde{Z}=\int\exp[-\phi_1(x,y)]{dxdy}/{\pi}$ with
\begin{eqnarray}
\phi_1(x,y)=\beta\omega (x^2+y^2)-\sum_{i=1}^N \ln 2\cosh[\beta\sqrt{\Delta_i^2+\frac{g_i^2(1+\lambda_i)^2x^2}{N}+\frac{g_i^2(1-\lambda_i)^2y^2}{N}}].
\end{eqnarray}
Let us change variables $u=\sqrt{{\omega}/{N\Delta}}x$ and $v=\sqrt{{\omega}/{N\Delta}}y$, where $\Delta$ is again a positive generalised mean of $\{\Delta_i\}$. Then the reduced free energy $f=-\ln\tilde{Z}/(\beta N\Delta)$ is well approximated by the value at its global minimum of the Landau potential
\begin{eqnarray}\label{phiu}
\phi(u,v)=u^2+v^2-\sum_{i=1}^N\frac{1}{\beta N\Delta}\ln 2\cosh[\beta\Delta\sqrt{\delta_i^2+\gamma_i^2(1+\lambda_i)^2u^2+\gamma_i^2(1-\lambda_i)^2v^2}],
\end{eqnarray}where $\delta_i={\Delta_i}/{\Delta}$ and $\gamma_i={g_i}/{\sqrt{\Delta\omega}}$.

Extrema of this function satisfy
\begin{eqnarray}\label{an1}
\frac{\partial \phi}{\partial u}=2u-\frac{1}{N}\sum_{i=1}^N \tanh[\beta \Delta\sqrt{\delta_i^2+\gamma_i^2(1+\lambda_i)^2u^2
+\gamma_i^2(1-\lambda_i)^2v^2}]\frac{\gamma_i^2(1+\lambda_i)^2u}{\sqrt{\delta_i^2+\gamma_i^2(1+\lambda_i)^2u^2
+\gamma_i^2(1-\lambda_i)^2v^2}}=0,\\
\frac{\partial \phi}{\partial v}=2v-\frac{1}{N}\sum_{i=1}^N \tanh[\beta \Delta\sqrt{\delta_i^2+\gamma_i^2(1+\lambda_i)^2u^2
+\gamma_i^2(1-\lambda_i)^2v^2}]\frac{\gamma_i^2(1-\lambda_i)^2v}{\sqrt{\delta_i^2+\gamma_i^2(1+\lambda_i)^2u^2
+\gamma_i^2(1-\lambda_i)^2v^2}}=0.\label{an2}
\end{eqnarray}
Clearly the origin $(u,v)=(0,0)$ is always an extremum, and the global minimum for $\gamma_i$ close to $0$. If $\lambda_i\neq0$, Eqs. \eqref{an1} and \eqref{an2} cannot be  simultaneously satisfied with both $u$ and $v$ different from 0. Let us first assume $u=0$. In this case, $v\neq0$ solutions to Eq. \eqref{an2} exist if
\begin{equation}\label{c1}
\frac{1}{N}\sum_{i=1}^N\tanh[\beta \Delta\delta_i]\frac{\gamma_i^2(1-\lambda_i)^2}{\delta_i}=
\sum_{i=1}^N\tanh[\beta \Delta_i]\frac{g_i^2(1-\lambda_i)^2}{N\omega\Delta_i}\geq 2\,.
\end{equation}
The transition is continuous.

Passing now to the case $v=0$, we see that a $u\neq0$ solution to Eq. \eqref{an1} exist if
\begin{equation}\label{c2}
\frac{1}{N}\sum_{i=1}^N\tanh[\beta \Delta\delta_i]\frac{\gamma_i^2(1+\lambda_i)^2}{\delta_i}=
\sum_{i=1}^N\tanh[\beta \Delta_i]\frac{g_i^2(1+\lambda_i)^2}{N\omega\Delta_i}\geq2,
\end{equation}
again leading to a continuous transition.

To illustrate this assertion, let us consider the homogeneous case in the limit $\beta\Delta\to\infty$. Criticality is achieved at $\gamma^2(1+\lambda)^2=2$ for $u\neq0$ solutions and at $\gamma^2(1-\lambda)^2=2$ for $v\neq0$. Close to criticality in the first case,  $u^2_{\mathrm{min}}\approx 2(|\gamma(1+\lambda)|-\sqrt{2})$.   

As particular cases of this Hamiltonian Eq. \eqref{an1}, notice that when all $\lambda_i=1$, then the inhomogeneous Dicke model is recovered. The SPT presents itself  as  $u$ type, with the condition for superradiance
\begin{equation}
\sum_{i=1}^N\tanh[\beta \Delta_i]\frac{2g_i^2}{N\omega\Delta_i}> 1.
\end{equation}
The phase diagram for homogeneous Dicke model is shown in Fig. \ref{fig1s}. Notice now that  $N=1$ with $\lambda=1$ corresponds to the Rabi model.

\begin{figure}[htbp]
\center
\resizebox{0.7\columnwidth}{!}{%
  \includegraphics{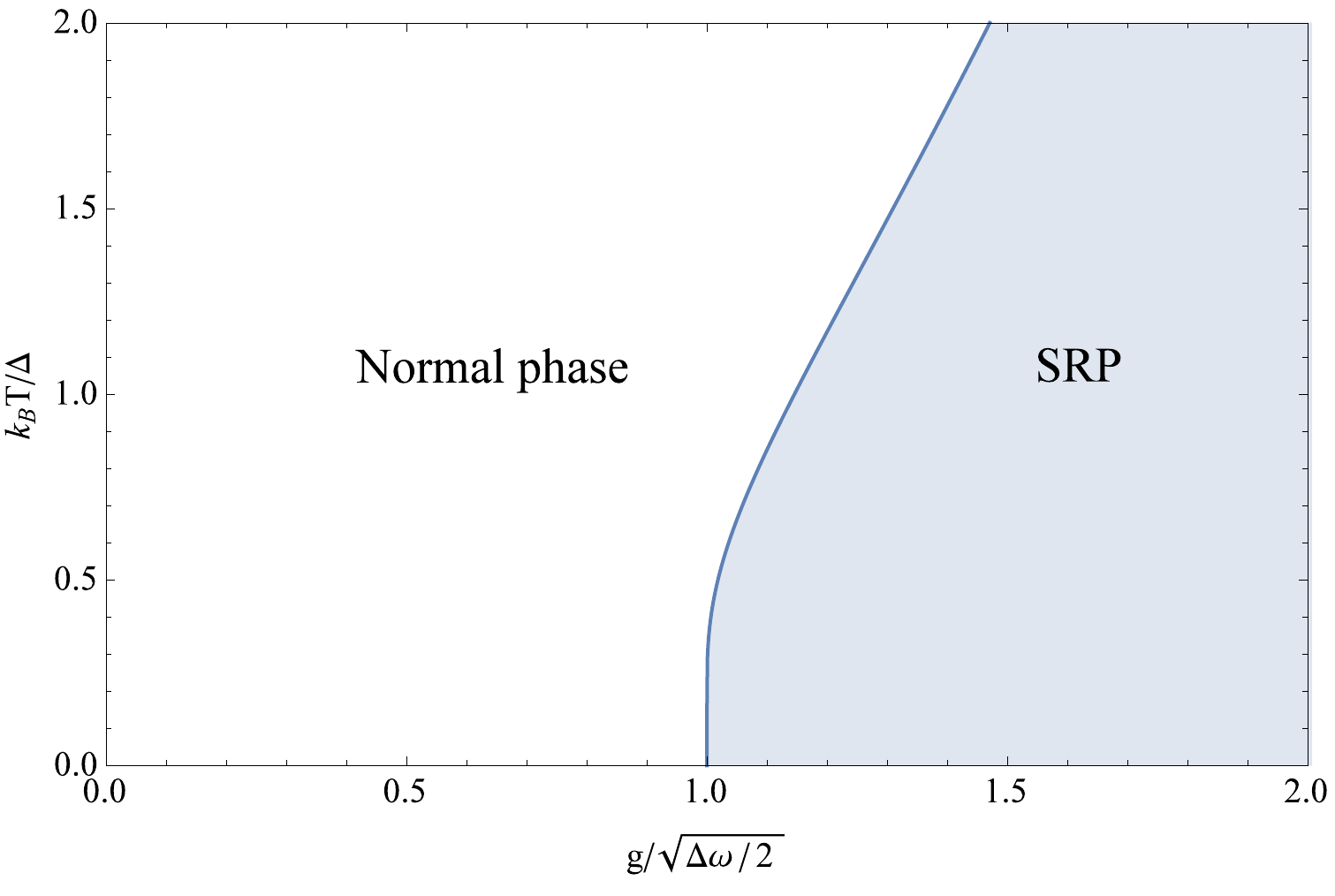}
}
\renewcommand\figurename{\textbf{Figure}}
\caption[2]{Phase diagram of the Dicke model in thermodynamic limit ($N\to\infty$).
}
\label{fig1s}
\end{figure}

For these special situations,  and for generic values of the parameters as well, the Hamiltonian presents a $\mathbb{Z}_2$ symmetry, with generator $\Pi=\exp\left\{i\pi\left[a^\dag a+\sum_{i=1}^N(1+\sigma_{iz})/2\right]\right\}$.
This is reflected by the two $u\leftrightarrow -u$ and  $v\leftrightarrow -v$ symmetries of $\phi(u,v)$. In fact it is a function of $u^2$ and $v^2$. It follows that the Hessian matrix at the origin is diagonal for all values of the parameters. The critical lines identified above correspond to a first sign change in the Hessian determinant at the origin, that is, when the origin in $(u,v)$ space goes from being a minimum to being a saddle point. We see here the usual behaviour of continuous transition Landau functions for $\mathbb{Z}_2$ symmetry.

The symmetry is enhanced if all  $\lambda_i$ are zero, in which case the Hamiltonian Eq. \eqref{an1} corresponds  to the inhomogeneous Tavis--Cummings model.
Eq. \eqref{phiu} reduces to
\begin{eqnarray}\label{phiu1}
\phi(u,v)=u^2+v^2-\sum_{i=1}^N\frac{1}{\beta N\Delta}\ln 2\cosh[\beta\Delta\sqrt{\delta_i^2+\gamma_i^2(u^2+v^2)}].
\end{eqnarray}
This function presents rotational symmetry in the $(u,v)$ plane, which is a reflection of the $U(1)$ symmetry of the inhomogeneous Tavis--Cummings model. The Hessian matrix at the origin is proportional to the identity,
with proportionality constant $2-(1/N)\sum_i\gamma_i^2\tanh(\beta \Delta_i)/\delta_i$. 
Criticality is determined by this quantity being zero, or, equivalently, by the Hessian determinant at origin being zero. That is, the system is in a superradiant phase if  \begin{equation}
\sum_{i=1}^N\tanh[\beta \Delta_i]\frac{g_i^2}{N\omega\Delta_i}> 2\,.
\end{equation}
The Landau function goes from having a global minimum at the origin to a Mexican hat like shape. Using polar coordinates, we can interpret the symmetry breaking as forcing the choice of a particular direction $\theta$, with $\theta$ the polar angle. All these directions are equivalent, so we expect the  gapless excitations.

Let us now study ground states, of particular relevance in the Quantum SPT
case. Namely,
once the minimum point has been determined, we can produce a variational estimate also for the ground state wavefunction. For concreteness, take the case   $\lambda_i=1$, which is the inhomogeneous  DQR Hamiltonian.
The effective mean field Hamiltonian reads
\begin{equation}
  \label{eq:inhomeff}
  H\left(u_{\mathrm{min}}\right)=N\Delta u_{\mathrm{min}}^2+\Delta\sum_{i=1}^N\left[2u_{\mathrm{min}}\gamma_i\sigma_{ix}+\delta_{i}\sigma_{iz}\right]\,,
\end{equation}
which is a sum of independent spin terms, readily diagonalizable. The ground state in spin space of $H\left(u_{\mathrm{min}}\right)$ is
\begin{equation}
  \label{eq:inhomspinground}
  |\psi_{sp}\rangle=\otimes_{i=1}^N
  \begin{pmatrix}
    \sin\theta_i\\ -\cos\theta_i
  \end{pmatrix}\,,
\end{equation}
where the angles are defined by
\begin{equation}
  \label{eq:1}
  \tan\theta_i=\frac{2u_{\mathrm{min}}\gamma_i}{\delta_i+\sqrt{\delta_i^2+4 u_{\mathrm{min}}^2\gamma_i^2}}\,.
\end{equation}
The average spin energy in the ground state, in units of $\Delta$, is therefore
\begin{equation}
  \label{eq:avspininhomground}
  \frac{1}{N}\left\langle\sum_{i=1}^N\sigma_{iz}\right\rangle= -\frac{1}{N}\sum_{i=1}^N\frac{\delta_i}{\sqrt{\delta_i^2+4\gamma_i^2u_{\mathrm{min}}}}\,.
\end{equation}
When the symmetry is broken, the ground state is determined by  a non-zero $u_{\mathrm{min}}$, resulting in the total ground state (both bosonic and spin spaces) $|gs\rangle=|\alpha_{\mathrm{min}}\rangle\otimes|\psi_{sp}\rangle$.

Outside of the generalised thermodynamic limit the mean field approach still provides us with a good approximation. The $\mathbb{Z}_2$ symmetry is however not broken, so the ground state, which presents odd parity, takes the approximate form
\begin{eqnarray}
  \label{eq:groundQRMSR}
  |\mathrm{GS}\rangle&=&\frac{1}{\sqrt{2}}\left[ |\alpha_{\mathrm{min}}\rangle\otimes_{i=1}^N
  \begin{pmatrix}
    \sin\theta_i\\ -\cos\theta_i
  \end{pmatrix}-|-\alpha_{\mathrm{min}}\rangle\otimes_{i=1}^N
  \begin{pmatrix}
    \sin\theta_i\\ \cos\theta_i
  \end{pmatrix}\right]\nonumber\\
  &=&\frac{1}{\sqrt{2}}\left[ (|\alpha_{\mathrm{min}}\rangle-|-\alpha_{\mathrm{min}}\rangle)|\psi_{sp+}\rangle
  +(|\alpha_{\mathrm{min}}\rangle+|-\alpha_{\mathrm{min}}\rangle)|\psi_{sp-}\rangle\right],
\end{eqnarray}
where $|\psi_{sp\pm}\rangle$ is the $|\pm\rangle$ eigenstate of $\prod_{i=1}^N\sigma_{iz}$.
Given the effectively zero overlap $\langle \alpha_{\mathrm{min}}|-\alpha_{\mathrm{min}}\rangle$, this state is normalized to be a ``Schr\"{o}dinger cat''-like state for finite $N$.
In single qubit Rabi case,  $u_{\mathrm{min}}=\frac{1}{\sqrt{2}}\sqrt{\frac{\gamma^2}{\gamma_c^2}-\frac{\gamma_c^2}{\gamma^2}}\,$, $\gamma_c=\sqrt{1/2}$, so
\begin{equation}
  \label{eq:angleforQRM}
  \tan\theta=\frac{2 u_{\mathrm{min}}\gamma}{1+\sqrt{1+4\gamma^2u_{\mathrm{min}}^2}}=\frac{\sqrt{\gamma^4-\gamma_c^4}}{\gamma^2+\gamma_c^2}=\sqrt{\frac{\gamma^2-\gamma_c^2}{\gamma^2+\gamma_c^2}} \,.
\end{equation}

Knowing the explicit ground state (in the variational approximation), we can portray many of its characteristics. For illustration, we consider the photon number distribution of the ground state of the quantum Rabi model ($N$=1) with $\Delta/\omega\gg1$.
 Now, the probability of measuring $n$ photons in this state while having  spin up, $P_{ne}$, will be zero for even $n$, and, similarly, the probability of odd photons with spin down will be zero. This results in
\begin{eqnarray}
  \label{eq:photonnumberdist}
  P_{2m+1,e}&=& 2\sin^2\theta\, e^{-|\alpha_{\mathrm{min}}|^2}\frac{|\alpha_{\mathrm{min}}|^{2(2m+1)}}{(2m+1)!}\,,\nonumber\\
  P_{2m,g}&=& 2\cos^2\theta\, e^{-|\alpha_{\mathrm{min}}|^2}\frac{|\alpha_{\mathrm{min}}|^{4m}}{(2m)!}\,.
\end{eqnarray}
We portray these expressions, together with those obtained by direct diagonalization, in Fig. \ref{fig2s}.

\begin{figure}[htbp]
\center
\resizebox{0.7\columnwidth}{!}{%
  \includegraphics{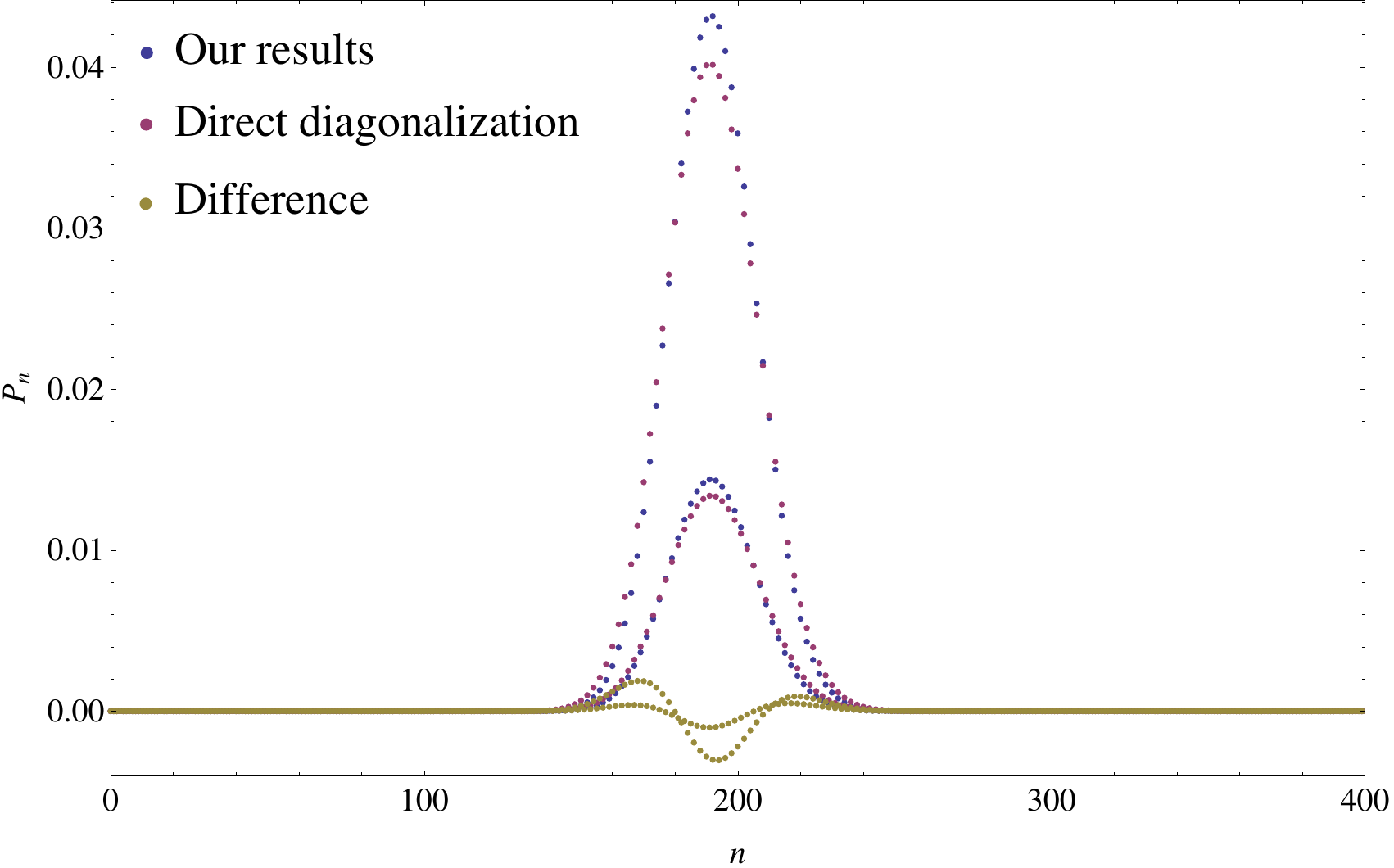}
}
\renewcommand\figurename{\textbf{Figure}}
\caption[2]{The ground state photon distribution of the Rabi model. $\Delta=16$, $\omega=1/16$, $g=1$, in arbitrary units, corresponding to $\gamma=1$, $\gamma_c=1/\sqrt{2}$, $C=\Delta/\omega=256$, $\langle a^\dagger a\rangle =Cu^2_{\mathrm{min}}=C|\alpha_{\mathrm{min}}|^2=192$. The red dots are obtained from  Eq. \eqref{eq:photonnumberdist}, while the blue dots are obtained by numerical diagonalization of the Hamiltonian. The yellow dots portray the difference.}
\label{fig2s}
\end{figure}

\section{No symmetry}
\label{sec:no-symmetry}

We have stressed in the analysis above that the $\mathbbm{Z}_2$ symmetry has guaranteed that the relevant $\phi(u)$ function is even. This has allowed us to establish the existence of a SPT and its second order character. Now we break this symmetry by adding a static bias. More concretely, we consider the Hamiltonian
\begin{eqnarray}\label{eq:dickeb}
  H&=&\omega a^\dagger a+\sum_{i=1}^{N} \frac{g_i}{\sqrt{N}} \sigma_{ix}(a+a^\dagger)+\sum_{i=1}^{N} \Delta_i \sigma_{iz}+\sum_{i=1}^{N}e_i\sigma_{ix}\\
  &=& \Delta\sum_{i=1}^N\left[\Omega\frac{a^\dag a}{C}+\gamma_i\sigma_{ix}\frac{a+a^\dag}{\sqrt{C}}+\delta_i\sigma_{iz}+\epsilon_i\sigma_{ix}\right]\,,
\end{eqnarray}
where, as before, $C$ will be the parameter or combination of parameters that give a macroscopic limit, while $\Omega$, $\gamma_i$, $\delta_i=\Delta_i/\Delta$ and $\epsilon_i=e_i/\Delta$ are finite, with $\Delta$ being some average of $\Delta_i$. Choosing $C=N\Delta/\omega$ we set $\Omega=1$, while  $\gamma_i=\frac{g_i}{\sqrt{\Delta_i\omega}}$.

This model presents no parity symmetry. Nonetheless, the analysis above will be applicable in that in both the thermodynamical and the classical oscillator limit the mean field approximation of the partition function will be adequate.
Following indeed the steps presented above we obtain the corresponding $\phi(u)$ function,
\begin{equation}
  \label{eq:phiinhomDickestaticbias}
  \phi(u)=u^2-\frac{1}{\beta\Delta N}\sum_{i=1}^N\ln\left[\cosh\left(\beta\Delta\sqrt{\delta_i^2+\left(\epsilon_i+2\gamma_i u\right)^2}\right)\right]\,.
\end{equation}
Clearly this is no longer an even function of $u$ if not all biases $\epsilon_i$ are zero, whence it follows that $u=0$ is not generically an extremum. For definiteness, let us concentrate on the homogenous case with finite $N$, in the classical oscillator limit. The 
Landau potential $\phi(u)$ is
\begin{equation}
  \label{eq:effectivephiQRMstatic}
  \phi(u)=u^2 -\sqrt{1+\left(\epsilon+2\gamma u\right)^2}\,.
  \end{equation}
 Its derivative is
 \begin{equation}
   \label{eq:phiQRMstaticderiv}
\phi'(u)=   2u-\frac{2\gamma(\epsilon+2\gamma u)}{\sqrt{1+\left(\epsilon+2\gamma u\right)^2}},.
\end{equation}
One readily sees that the origin is never an extremum for generic $\epsilon$ and $\gamma$: The system always presents a macroscopic photon number. 

\section{Dicke model with one and two-photon terms}
\label{sec:dicke-model-with}
A relevant question is whether models with dipolar coupling to \emph{two} photons can also present superradiance. As we shall see, in general the parity symmetry we have so much relied upon will not be available, so any transition, if present, cannot be continuous. In order to assess this point, we will, 
as before,  rewrite the Hamiltonian Eq. \eqref{eq:Dicke12phtFirstHam} in terms of a set of finite parameters $\Omega$, $\gamma$ and $\gamma'$ and a macroscopicity parameter $C=N\Delta/\omega$. That is,
\begin{equation}
  \label{eq:Dicke12phtHam}
  H=\Delta\sum_{i=1}^N\left[\frac{a^\dag a}{C}+\gamma\frac{a+a^\dag}{\sqrt{C}}\sigma_{ix}+\gamma'\frac{a^2+\left(a^\dag\right)^2}{C}\sigma_{ix}+\sigma_{iz}\right]\,,
\end{equation}
where $\gamma$=$g/\sqrt{\Delta\omega}$, $\gamma'=g'/\omega$.
Following the by now well established routine, we obtain the reduced free energy $F/(N\Delta)$ as the minimum of a potential, now in two variables,
\begin{equation}
  \label{eq:twovarthermpot} \phi(u,v)=u^2+v^2-\frac{1}{\beta\Delta}\ln\left[2\cosh\left(\beta\Delta\sqrt{1+\left[2\gamma u+2\gamma'\left(u^2-v^2\right)\right]^2}\right)\right]\,.
\end{equation}
The stability of the system requires that the minima be located at a finite point in $(u,v)$ space. Using polar coordinates, the dominant large radius behaviour of the potential is $r^2\left(1-2\gamma'\cos(2\theta)\right)$. Thus, stability requires $|\gamma'|<1/2$, or, in the original parameters, $|g'|<\omega/2$. This result has been obtained in different forms in the literature, and the limiting value corresponds to the spectral collapse due to the two-photon term.

This model has no symmetry for nonzero $\gamma'$ and $\gamma$. Nonetheless,
$\phi(u,v)$ is an even function of $v$ for all values of the parameters. As is only to be expected, there are some regions in parameter space for which a symmetry does exist. Namely, if $\gamma'=0$ we recover the Dicke case with $\mathbbm{Z}_2$ symmetry, and this fact is reflected in $\phi$ being an even function of $u$ as well. If $\gamma=0$ we have a $\mathbbm{Z}_4$ symmetry arising from $u\leftrightarrow v$ exchange symmetry and parity.
Now we search for extrema of $\phi(u,v)$ on computing the its derivative
 \begin{eqnarray}
 \frac{\partial \phi(u,v)}{\partial u}=2u-\tanh[\beta\Delta\sqrt{1+\left[2\gamma u+2\gamma'\left(u^2-v^2\right)\right]^2}]\frac{(2\gamma+4\gamma'u)\left[2\gamma u+2\gamma'\left(u^2-v^2\right)\right]}{\sqrt{1+\left[2\gamma u+2\gamma'\left(u^2-v^2\right)\right]^2}},\label{uv1}\\
 \frac{\partial \phi(u,v)}{\partial v}=2v+\tanh[\beta\Delta\sqrt{1+\left[2\gamma u+2\gamma'\left(u^2-v^2\right)\right]^2}]\frac{4\gamma'v\left[2\gamma u+2\gamma'\left(u^2-v^2\right)\right]}{\sqrt{1+\left[2\gamma u+2\gamma'\left(u^2-v^2\right)\right]^2}}.\label{uv2}
 \end{eqnarray}
From Eq.\eqref{uv2}, one immediately sees that $\partial_v\phi(u,0)$ is identically zero. In the alternative, one obtains a line in the $(u,v)$ plane. On computing the first equation on this line, one obtains a linear equation for $u$. On substituting back into the line to determine critical points, one sees that there are no real solutions for $v$, as  long as  the stability condition is satisfied. It follows that all the extrema are located on the $v=0$ line. We thus have to analyze the extrema of
\begin{equation}
  \label{eq:reducedphiDicke12pht}
  \tilde\phi(u)=u^2 -\frac{1}{\beta\Delta}\ln\left[2\cosh\left(\beta\Delta\sqrt{1+\left[2\gamma u+2\gamma'u^2\right]^2}\right)\right]\,.
\end{equation}
Clearly $u=0$ is the global minimum when $\gamma$ and $\gamma'\to0$.

In the limit $\beta\Delta\to\infty$ the crucial equations become algebraic. In particular, in this limit a nonzero solution to $\phi(u)=\phi(0)$ will exist when $2\gamma^2+4\gamma'^2\geq1$, and in fact $\phi^\prime(u)=0$ at criticality, $2\gamma^2+4\gamma'^2\geq1$. We thus establish the existence of a first order superradiant transition with the aforementioned critical line.

We illustrate our results in the main text by diagonalizing the Hamiltonian with $N=1$ and ${\Delta}/{\omega}=200$.  We portray the  expected first order quantum phase transition  in Fig. \ref{fig4s}(a) and \ref{fig4s}(b). The  crossing between ground and first excited state is necessarily avoided with finite parameters. Nonetheless, the change from zero to macroscopic  photon number is extremely sharp, as expected from the analytic results for  the limiting case of $\Delta/\omega\to\infty$.
\begin{figure}[htbp]
\center
\resizebox{1\columnwidth}{!}{%
  \includegraphics{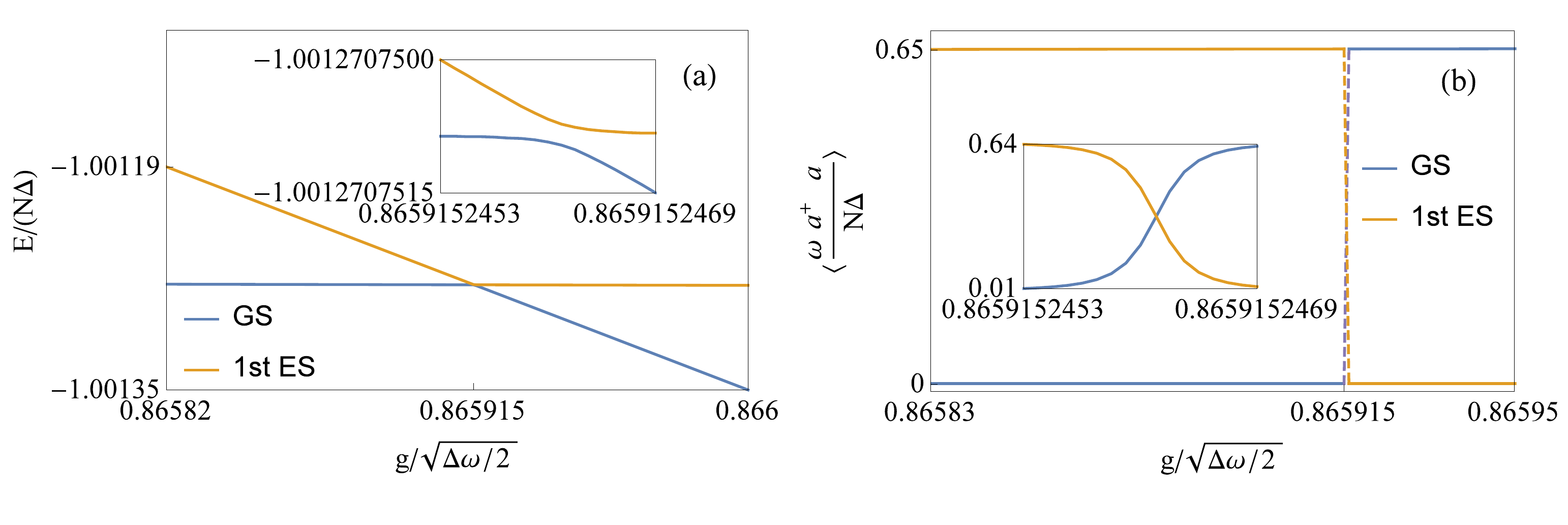}
}
\renewcommand\figurename{\textbf{Figure}}
\caption[2]{(a)Energies of the ground state (GS) and 1st
excited state (ES) obtained by direct diagonalization of the Hamiltonian Eq. \eqref{eq:Dicke12phtFirstHam} with $N=1$, $\Delta/N\omega=200$, $g^\prime/\omega=0.25$, $\frac{g_c}{\sqrt{\Delta\omega/2}}=0.866$. $\langle \frac{\omega a^\dagger a}{N\Delta}\rangle_c=2/3$. (b)The photon number of the GS and 1st ES in (a). $\langle a^\dagger a\rangle_c=400/3$. The numerical results match the exact  analysis for $N\Delta/\omega\to\infty$.
\label{fig4s}}
\end{figure}

\section{General structure of the phase diagram}
\label{sec:gener-struct-phase}

The single mode models we are considering have the general structure
\begin{equation}
  \label{eq:genham}
  \frac{1}{\omega}H= a^\dag a+ A + B a^\dag +B^\dag a + D \left(a^\dag\right)^2+D^\dag a^2\,,
\end{equation}
with $A$, $B$ and $D$ operators that act on a finite dimensional Hilbert space $\mathfrak{H}$. These operators depend continuously on the (reduced) parameters that define the phase space of interest. We have shown above that, as long as $\|D^\dag D\|<\omega^2/4$, the mean field approximation will be valid for (generalised) thermodynamic limits, and that the phase structure will be determined by a Landau potential of the form
\begin{equation}
  \label{eq:genphi}
  \phi(u,v)=u^2+v^2+\lambda(u,v)\,,
\end{equation}
or possibly reduced to one variable $u$.
The symmetries of $\phi(u,v)$ will be inherited from those of the Hamiltonian. This function is continuous and differentiable in the situations we study. Furthermore, it is also continuous and even differentiable with respect to the parameters of the model.  In all cases of interest, the dominant term for large $u^2+v^2$ will be quadratic in these variables. Additionally, in most cases  the origin will be an extremum of the Landau potential. Not so, let it be noted, when a bias is introduced. Given these two considerations, namely that for large order parameter the Landau potential grows quadratically, and that the origin is an extremum for all values of the parameters, it follows that there are two alternatives: either the origin is the unique global minimum or there exists at least one non zero point $(u_*,v_*)$ such that $\phi(u_*,v_*)=\phi(0,0)$. This special point (or points) will depend on the reduced parameters.

If indeed the system is in the superradiant region of phase space, the potential will have a global minimum away from the origin. The location of the minimum will also depend on the parameters, non analytically at the phase transition.

In the cases of interest, there is a set of \emph{dipolar} couplings $\vec{\gamma}$, such that 
the potential can be written as
\begin{equation}
  \label{eq:dipolarpotential}
  \phi(u)=u^2+\mu(u\vec{\gamma})\,,
\end{equation}
with $\mu$ a function of several variables, $\mu(\vec{x})$. Notice that this indeed applicable to the Dicke model with both one and two photon terms presented in Eq. \eqref{eq:Dicke12phtHam} of Section \ref{sec:dicke-model-with}, with $\vec{\gamma}=(\gamma,\sqrt{\gamma'})$.  We readily write $\partial_u\phi=2u+\vec{\gamma}\cdot\nabla \mu$, where we denote with $\nabla$ the gradient with respect to the variables of function $\mu$.
Since $\phi$ depends on the dipolar couplings, we can also consider its gradient in dipolar coupling space, $\nabla_\gamma$. Clearly we have $\nabla_\gamma\phi=u \nabla\mu$.  Combining these two results, we conclude
\begin{equation}
  \label{eq:genphiugamma}
  u\partial_u\phi=2 u^2+\vec{\gamma}\cdot\nabla_\gamma\phi\,.
\end{equation}
We see that if there is an extremum of $\phi$ away from the origin, $\vec{\gamma}\cdot\nabla_\gamma\phi$ will be negative at that point. Hence, the radial derivative in dipolar coupling space is negative at that point, and if the nonzero extremum is a minimum, an increase in coupling it will deepen. Thus, once in a superradiant phase, a radial increase in the dipolar couplings will move us further into the superradiant phase. 

\section{Two-qubit Rabi model with Heisenberg interaction}
\label{sec:two-qubit-rabi}

As we have stated, in the family of models under consideration going radially in dipolar coupling space one moves deeper into the superradiant phase, once that superradiance has been achieved. It is not the case, however, that there is isotropy in dipolar couplings. To illustrate this point, we have mentioned in the main text, Eq.~\eqref{eq:2spinXYZ}, the  following model that we restate here,
\begin{equation}
  \label{eq:2spinXYZSM}
H=\omega a^\dagger a+\sum_{j=1,2}[g_j\sigma_{jx}(a+a^\dagger)+\Delta_j \sigma_{jz}]+\sum_{\alpha=x,y,x}J^{(\alpha)}\sigma_{1\alpha}\sigma_{2\alpha}\,.
\end{equation}
This model presents generically a $\mathbb{Z}_2$ symmetry, with operator $\Pi=\exp\left\{i\pi\left[a^\dag a+\sum_{i=1}^2\left(1+\sigma_{iz}\right)/2\right]\right\}$.

Let us define $\Delta$ as some average of the spin energies $\Delta_i$, be it $\Delta=\sum_i\Delta_i/N$ if not zero,  or $\Delta_i=\left(\sum_i\Delta_i^2/N\right)^{1/2}$, or some other generalised mean. We now  define $\gamma_j=g_j/\sqrt{\Delta\omega}$, $\delta_j=\Delta_j/\Delta$ and $\epsilon_\alpha=J^{(\alpha)}/\Delta$. After following the same analysis as before for other models, we arrrive at an effective mean field spin hamiltonian
\begin{equation}
  \label{eq:XYZmeanfieldspinham}
  h(u)=\sum_{j=1}^{2}\left(2\gamma_j u \sigma_{jx}+\delta_j \sigma_{jz}\right)+\sum_\alpha \epsilon_\alpha \sigma_{1\alpha}\sigma_{2\alpha}\,.
\end{equation}
We study this model in the limit $\beta \Delta\to\infty$. In this case the phase structure is controlled by the Landau potential $\phi(u)=u^2+\lambda(u)$, with $\lambda(u)$ the smallest eigenvalue of $h(u)$.

In this case, as we shall see, there is generically an anisotropy in the space of dipolar couplings due to the presence of the spin interactions and different spin energies, together with inhomogeneous dipolar couplings. In fact, in the limiting case $\epsilon_\alpha\to0$, which we have studied earlier, it is trivial to determine that $\lambda(u)=-\sum_j\sqrt{\delta_j^2+4\gamma_j^2u^2}$.
This provides us with the critical line
\begin{equation}
  \label{eq:critical2}
  2\sum_j\frac{\gamma_j^2}{\left|\delta_j\right|}=1\,.
  \end{equation}
  This critical line is an ellipse, more or less elongated depending on the inhomogeneity of $\delta_j$. It is important to notice that the actual shape of the critical line can be computed perturbatively in this case. To understand this statement, consider the case in which the spin energies are the same and positive, $\delta_1=\delta_2=\delta>0$. The smallest eigenvalue of $h(0)$ will be $-2\delta$. Computing $\lambda(u)$ perturbatively to second order, the corresponding eigenvalue reads $-2\delta-2(\gamma_1^2+\gamma_2^2) u^2/\delta $.
  This comes about because in these cases the transition is second order in all directions.

  In this general model, however, the possibility exists of first order transitions. To show that this is indeed the case, we consider the homogeneous model, in which $\gamma_1=\gamma_2=\gamma$ and $\delta_1=\delta_2=\delta$. In this case we rewrite the effective spin hamiltonian as
  \begin{equation}
    \label{eq:hommodel}
    h(u)= 4\gamma u S_x+2\delta S_z+2\sum_\alpha \epsilon_\alpha S_\alpha^2-\sum_\alpha\epsilon_\alpha\,,
  \end{equation}
where $\mathbf{S}=(\boldsymbol{\sigma}_1+\boldsymbol{\sigma}_2)/2$.
In the singlet subspace the eigenvalue is $-\sum_\alpha\epsilon_\alpha$. This eigenvalue is independent of $u$. Thus, consider the situation in which there is a point $u_*$ at which the smallest eigenvalue of $h(u)$ in the triplet space crosses the constant eigenvalue in the singlet subspace. The Landau potential, for $u<u_*$, is of the form $u^2-\sum_\alpha \epsilon_\alpha$, while for $u>u_*$ there is an additional $u$ dependence, which can introduce a new minimum at finite distance from the origin. Consider, for example, a situation in which $\epsilon_y=0$, $\epsilon_x>0$ and $\epsilon_z>\sqrt{\epsilon_x^2+4\delta^2}-\epsilon_x>0$. In this case the smallest eigenvalue of $h(u)$ close to $u=0$ is the singlet space one. As $u$ grows, the dominant term of $h(u)$ will be the first one, and the smallest eigenvalue of $h(u)$ for large $u$ will be $-4\gamma u$. Clearly at some point the smallest eigenvalue of the triplet subspace will cross the constant singlet eigenvalue. Furthermore, from that point on the derivative of the Landau potential will behave as $u^2-4\gamma u$ plus correction terms, thus having a minimum with smaller value than that at the origin. This means that the transition can in this case be a first order one.

This analysis suggests rewriting the effective spin hamiltonian as
\begin{eqnarray*}
  \label{eq:XYZsinglettriplet}
  h(u)&=&2\left(\gamma_1+\gamma_2\right) u S_x+\left(\delta_1+\delta_2\right) S_z+2\sum_\alpha \epsilon_\alpha S_\alpha^2\\
      &&-\sum_\alpha\epsilon_\alpha\\
  & &+\left(\gamma_1-\gamma_2\right)u\left(\sigma_{1x}-\sigma_{2x}\right)+\frac{1}{2}\left(\delta_1-\delta_2\right)\left(\sigma_{1z}-\sigma_{2z}\right)\,.
\end{eqnarray*}
In this form, the first line acts only on the triplet space, the second line is proportional to the identity, and the third line connects the singlet and the triplet spaces. At large $u>0$, the smallest eigenvalue will be $-\left(|\gamma_1|+|\gamma_2|\right) u$. One sees that the transition can be first order if this behaviour sets in before being overwhelmed by the $u^2$ term and the behaviour close to the origin of the smallest eigenvalue is very weakly dependent on $u$.

We have thus shown, qualitatively, that anisotropy is present in dipole coupling space, and that both first and second order transitions are feasible for some range of parameters.

Let us consider now an specific case for which a complete analysis can be easily provided; namely, identical spins ($\gamma_1=\gamma_2=\gamma$, $\delta_1=\delta_2=\delta$) with isotropic spin--spin coupling ($\epsilon_x=\epsilon_y=\epsilon_z=\epsilon$). Under the condition $\epsilon>|\delta|/2$ the Landau potential reads
\begin{equation}
  \label{eq:landauisotropic}
  \phi(u)=
  \begin{cases}
    u^2-3\epsilon& \mbox{if } |u|\leq\frac{1}{2|\gamma|}\sqrt{4\epsilon^2-\delta^2}\,,\\
    u^2+\epsilon-2\sqrt{\delta^2+4\gamma^2u^2}&\mbox{if }|u|\geq\frac{1}{2|\gamma|}\sqrt{4\epsilon^2-\delta^2}\,.
  \end{cases}
\end{equation}
Differentiating the second expression, we see that its minimum (other than zero) would be located at $|u_{\mathrm{m}}|=2|\gamma|\sqrt{1-\delta^2/16\gamma^2}$. This is the minimum, corresponding to a superradiant phase, if $16\gamma^4>4\epsilon^2>\delta^2$. Once $\epsilon$ and $\delta$ are fixed (and the condition $\epsilon>|\delta|/2$ holds), the critical point for a first order transition is given by $\gamma_c^2=\epsilon/2$.

On the other hand, if $\epsilon<|\delta|/2$ the smallest eigenvalue is always $\epsilon-2\sqrt{\delta^2+4\gamma^2 u^2}$, for all values of $u$. There will be a continuous transition with critical dipolar coupling $\gamma_c^2=|\delta|/4$.
The first and second order critical surfaces in the three dimensional space of $\gamma$, $\epsilon$ and $\delta$ have the common line $\gamma^2=\epsilon/2=|\delta|/4$.

In order to go forward and provide a more quantitative assessment, it is convenient to remember that the Landau potential will not have a unique global minimum at the origin $u=0$ if there exists a finite, nonzero value of $u$ such that $\phi(u)=\phi(0)$ holds. In the case at hand, this corresponds to the equation $\lambda(u)=\lambda(0)-u^2$. Since the determination of the smallest eigenvalue for generic parameters and order parameter $u$ is rather involved as a first step, and the computation of the derivative of the Landau potential hinges on this first computation, we sidestep the process in some cases by inserting the equation $\lambda(u)=\lambda(0)-u^2$ in the secular equation $\mathrm{Det}\left[h(u)-\lambda(u)\mathbbm{1}\right]$. This technique provides us with an analytical tool that can be also implemented numerically. We portray a number of cases in Figures \ref{fig5s}, \ref{fig6s} and \ref{fig7s}. To be more explicit, the substitution we have mentioned leads us to a cubic equation in the variable $u^2$. We are looking for real positive roots of the corresponding cubic polynomial. In particular we are looking for the existence of a real positive double root in order to identify criticality. This entails the discriminant of the cubic being zero. One also must look at the value of the polynomial at $u^2=0$, to assess whether the number of positive zeros (with multiplicities) is even or odd. Given some parameters for the spin energies and spin--spin couplings, this provides us with analytic equations for the dipolar couplings, and the result can be checked by numerical diagonalization.

This analysis provides us with a specific check on the general phase structure on the space of dipolar couplings presented above in Section \ref{sec:gener-struct-phase}. Indeed, as shown in the main text, the region close to the origin of dipolar couplings is in normal phase, and one goes radially into a superradiant one in the cases portrayed there.

     \begin{figure}[htbp]
\center
\resizebox{0.9\columnwidth}{!}{%
  \includegraphics{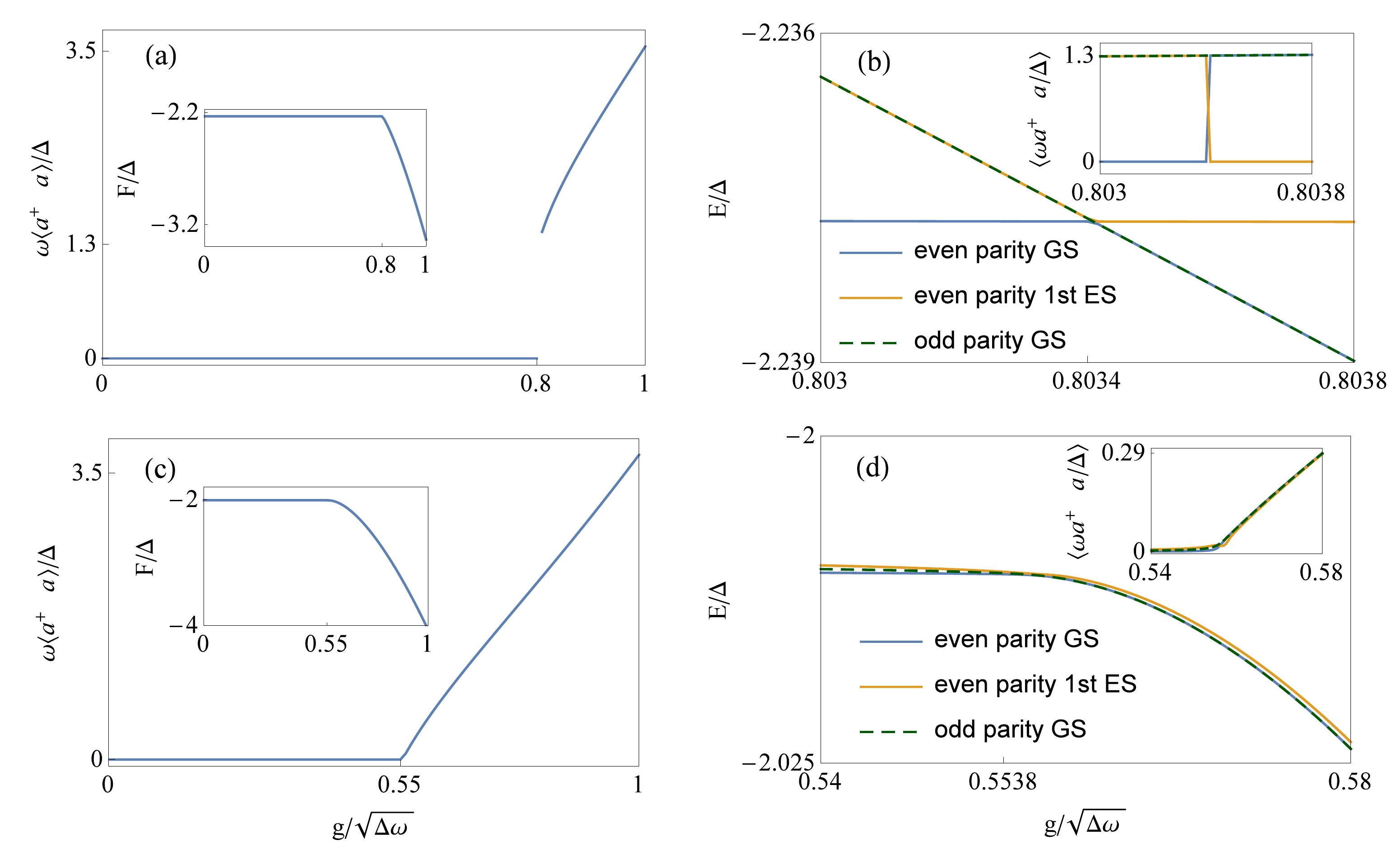}
}
\renewcommand\figurename{\textbf{Figure}}
\caption[2]{Left panel: The average rescaled photon number and reduced free energy of the identical two-qubit Rabi model with qubit dipole interaction at any finite temperature $\omega/k_B T\neq 0$ obtained by finding the global minimum of $\phi(u)$ numerically. Right panel: The lowest eigenenergy levels in each parity subspace and their average photon numbers obtained by diagonalizing the Hamiltonian numerically. (a)$\epsilon_x=1$. $\gamma_c=0.803$, $\langle \frac{\omega a^\dagger a}{\Delta}\rangle_c\sim1.308$. (b)$\Delta/\omega=C=200$, $\epsilon_x=1$, $\langle a^\dagger a\rangle_c\sim261.6$. (c)$\epsilon_x=0.2$, $\gamma_c=0.55387$.(d)$\Delta/\omega=800$,~$\epsilon_x=0.2$. All the critical properties are obtained analytically, coinciding well with the numerical results.
\label{fig5s}}
\end{figure}
\begin{figure}[htbp]
\center
\resizebox{1\columnwidth}{!}{%
  \includegraphics{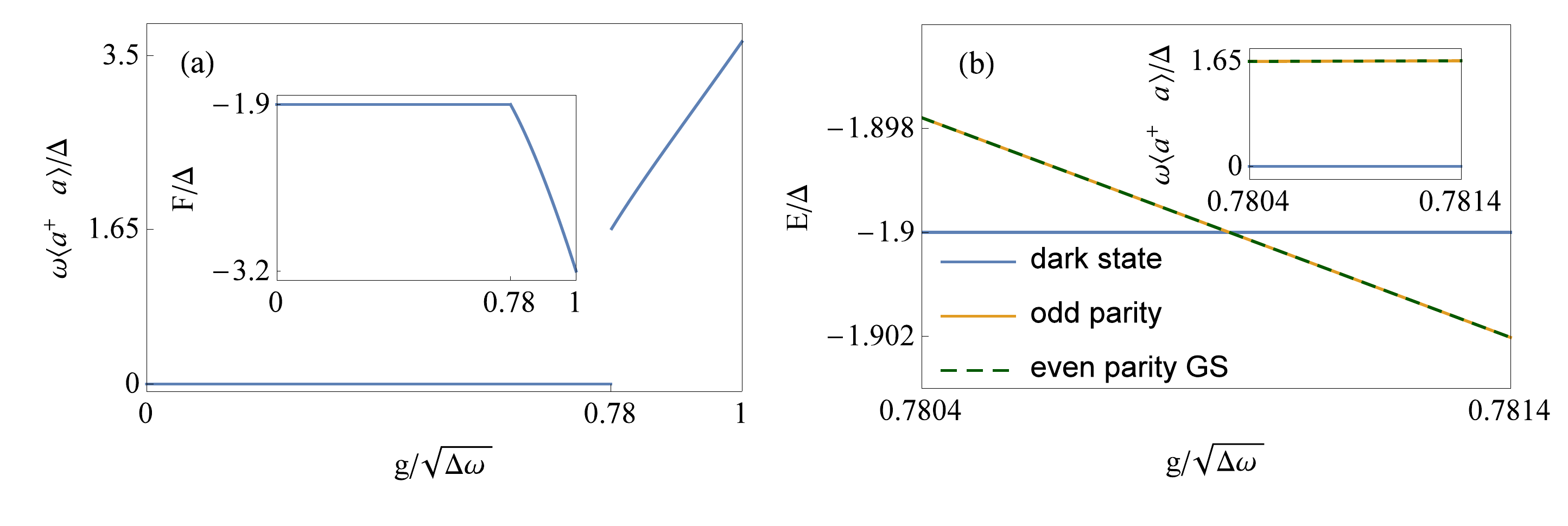}
}
\renewcommand\figurename{\textbf{Figure}}
\caption[2]{Left panel: Identical two-qubit Rabi model with XYZ Heisenberg interaction $\epsilon_x=1.1$, $\epsilon_y=0.3$, $\epsilon_z=0.5$ at any finite temperature $\omega/k_B T\neq0$.(a)The average rescaled photon number and reduced free energy obtained by finding the global minimum of $\phi(u)$ numerically. $\gamma_c=0.781063$ and $u^2_c=1.65378$.(f)Lowest energy levels obtained by direct diagonaliztion of the Hamiltonian for $\frac{\Delta}{\omega}=200$. $\langle a^\dagger a\rangle_c\sim330$. All the critical properties are obtained analytically, coinciding well with the numerical results.
\label{fig6s}}
\end{figure}
 \begin{figure}[htbp]
\center
\resizebox{1\columnwidth}{!}{%
  \includegraphics{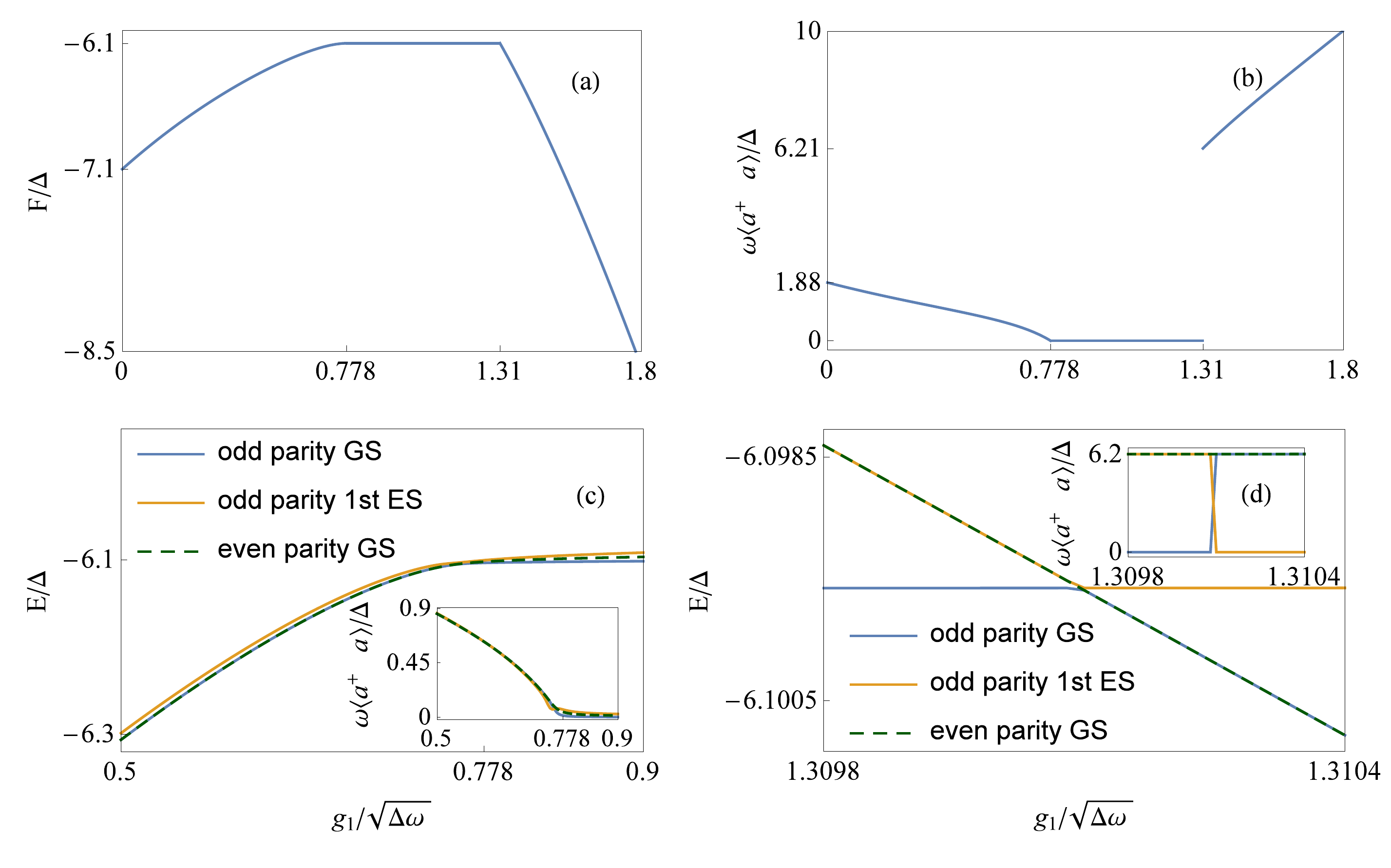}
}
\renewcommand\figurename{\textbf{Figure}}
\caption[2]{Nonidentical two-qubit Rabi model with XYZ spin-spin interactions. $\Delta_1/\Delta=3$, $\Delta_2/\Delta=2$, $J^x/\Delta=3$, $J^y/\Delta=2$, $J^z/\Delta=1$, $\frac{g_2}{\sqrt{\Delta\omega}}=1.5$, at any finite temperature $\omega/k_B T\neq0$.(a)Reduced free energy. (b)Rescaled photon number. The 2nd order SPT happens at $\frac{g_{1a}}{\sqrt{\Delta\omega}}=0.778$, while the 1st order SPT takes place at $\frac{g_{1b}}{\sqrt{\Delta\omega}}=1.31$, $\langle \frac{\omega a^\dagger a}{\Delta}\rangle_c\sim6.21$.  (c) and (d)Lowest energy levels obtain by direct diagonalizing the Hamiltonian at $\frac{\Delta}{\omega}=100$. $\langle a^\dagger a\rangle_c\sim6.20$.
\label{fig7s}}
\end{figure}
  \section{Multi-mode Dicke-QR model}
According to our analysis, the partition function can be written as
\begin{eqnarray}
Z&=&\int\ldots\int \mathrm{Tr}_{sp}\exp\{-\beta[\sum_\nu\omega_\nu(x_\nu^2+y_\nu^2)+\frac{\sum_{i,\nu}(2g_\nu x_\nu)\sigma_{ix}}{\sqrt{N}}+\sum_{i}
\Delta_i\sigma_{iz}]\}\frac{dx_1dy_1\ldots dx_\nu dy_\nu}{\pi^M}.
\end{eqnarray}
After tracing out the qubit and $y$ part, we obtain
\begin{eqnarray}
Z=\sqrt\frac{1}{\pi^M\beta^M\omega_1\omega_2\ldots\omega_M}\int \exp\{-\beta\sum_{\nu}\omega_\nu x_\nu^2
+\sum_{\i}\ln\left[2\cosh[\beta\Delta\sqrt{\delta_i^2+\frac{4(\sum_\nu g_\nu x_\nu)^2}{N\Delta^2}}]\right]\}dx_1\ldots dx_M,
\end{eqnarray}
Making the transformation $x_\nu\rightarrow \sqrt{\frac{N\Delta}{\omega_\nu}}u_\nu$, we obtain $Z=\mathbf{N^{\prime} }\int \exp\{-\beta N \Delta \phi(u_\nu)\}du_\nu$, with
\begin{eqnarray}
\phi(u_\nu)=\sum_\nu u_\nu^2-\frac{1}{\beta N\Delta} \sum_{i}\ln\left[2\cosh{{\beta \Delta\sqrt{\delta_i^2+4\big{(}\sum_\nu \gamma_\nu u_{\nu}}\big{)}^2}}]\right],
\end{eqnarray}
where $\gamma_\nu=\frac{g_{\nu}}{\sqrt{\omega_\nu\Delta}}$, $\delta_i=\Delta_i/\Delta$, $\Delta$ is the average of $\Delta_i$. For finite temperature $\beta\gg\frac{1}{N\Delta}$,
$\beta N \Delta$ is infinite, so according to Laplace's method, the reduced free energy $\frac{F}{N\Delta}$ is determined by the global minimum of $\phi(u_m)$, $m=1,2,\ldots,M$, which should satisfy
\begin{eqnarray}\label{mm1}
\frac{\partial\phi}{\partial u_m}=u_m-\frac{1}{N}\sum_i\tanh[\beta\Delta\sqrt{\delta_i^2+4\big{(}\sum_\nu \gamma_\nu u_{\nu}\big{)}^2}]\frac{2\sum_\nu (\gamma_\nu u_{\nu})\gamma_m}{\sqrt{\delta_i^2+4\big{(}\sum_\nu \gamma_\nu u_{\nu}\big{)}^2}}=0,
\end{eqnarray}
so $u_\nu=\frac{\gamma_\nu u_1 }{\gamma_1}$. Substituting this into Eq. \eqref{mm1} for $m=1$, and defining $\gamma^2=\sum_\nu \gamma_\nu^2$, we obtain
\begin{eqnarray} \label{multi}
&u_1\left(1-\frac{1}{N}\sum_i\tanh[\beta\Delta\sqrt{\delta_i^2+4\gamma^4u_1^2/\gamma_1^2}]\frac{2\gamma^2}{\sqrt{\delta_i^2+4\gamma^4 u_1^2/\gamma_1^2}}\right)=0.
\end{eqnarray}
One solution is $u_1=0$ and hence $u_\nu=0$, which is clearly global minimum when $g\sim0$. $\gamma$-dependent solution exists for
\begin{equation}
\gamma^2\geq\frac{N}{2\sum_i(\tanh[\beta\Delta_i]/\delta_i)}.
\end{equation}
At the critical $\gamma_c$, $u_\nu(\gamma_c)=0$, so $f(u_\nu(\gamma_c))=f(0)$, but as $\gamma_\nu$ increases, all nonzero $\phi(u_\nu)$ are decreasing, so the global minimum should be this $\gamma$-dependent extreme. For identical qubit $\Delta_i=\Delta$, it is easy to obtain the critical $\gamma_c^2=1/2$ and $\frac{\omega_\nu a^\dag a}{N\Delta}=u_\nu^2=\frac{\gamma_\nu^2}{4}(\frac{1}{\gamma_c^4}-\frac{1}{\gamma^4})$ for SRP in classical oscillator or zero temperature case ($\beta\Delta\to\infty$) from Eq. \eqref{multi}.

\end{widetext}


\begin{thebibliography}{10}

\bibitem{PhysRev.93.99}
R.~H. Dicke, Phys. Rev. \textbf{93}, 99 (1954).

\bibitem{hepp}K. Hepp and E. H. Lieb, Ann. Phys. \textbf{76}, 360 (1973).
\bibitem{wang}Y. K. Wang and F. T. Hioe, Phys. Rev. A \textbf{7}, 831 (1973).
\bibitem{PhysRev.170.379}
M.~Tavis and F.~W. Cummings.
 Phys. Rev. \textbf{170}, 379 (1968).

\bibitem{PhysRevA.8.1440}
F.~T. Hioe, Phys. Rev. A \textbf{8}, 1440 (1973).
\bibitem{PhysRevA.8.2517}
K.~Hepp and E.~H. Lieb,  Phys. Rev. A  \textbf{8}, 2517 (1973).
\bibitem{zhang}X.-Y. Chen and Y.-Y. Zhang, Phys. Rev. A \textbf{97}, 053821 (2018).
\bibitem{PhysRevA.95.053854}
L.~Garbe, I.~L. Egusquiza, E.~Solano, C.~Ciuti, T.~Coudreau, P.~Milman, and
  S.~Felicetti,
Phys. Rev. A \textbf{95}, 053854  (2017).
\bibitem{guo}X. Y. Guo, Z. Z. Ren, and Z. M. Chi, J. Opt. Soc. Am. B \textbf{5}, 1245 (2011).
\bibitem{li}C. F. Lee and N. F. Johnson, Phys. Rev. Lett. \textbf{93}, 083001 (2004).
\bibitem{li1}T. C. Jarrett, C. F. Lee and N. F. Johnson, Phys. Rev. B \textbf{74}, 121301(R) (2006).
\bibitem{GJC18} R. Guti\'errez-J\'auregui and H. J. Carmichael, Phys. Rev. A (R) \textbf{98}, 023804 (2018).


\bibitem{Bakemeier}L. Bakemeier, A. Alvermann and H. Fehske, Phys. Rev. A \textbf{85}, 043821 (2012).
\bibitem{PhysRevA.87.013826}
S.~Ashhab.  Phys. Rev. A \textbf{87} 013826 (2013).
\bibitem{hwang1}M.-J. Hwang, R. Puebla, and M. B. Plenio, Phys. Rev. Lett. \textbf{115},
180404 (2015).
\bibitem{hwang2}M.-J. Hwang and M. B. Plenio, Phys. Rev. Lett. \textbf{117}, 123602 (2016).

\bibitem{liu1}M. X. Liu, S. Chesi, Z.-J, Ying, X. S. Chen, H.-G. Luo and H.-Q. Lin, Phys. Rev. Lett. \textbf{119}, 220601 (2017).
\bibitem{lv}L.-T. Shen, Z.-B. Yang, H.-Z. Wu, and S.-B. Zheng, Phys. Rev. A \textbf{95}, 013819 (2017).
\bibitem{larson}J. Larson and E. K Irish 2017 J. Phys. A: Math. Theor. \textbf{50}, 174002 (2017).

\bibitem{sachdev_2011}
S.~Sachdev,  {\em Quantum Phase Transitions}, Cambridge University Press, 2 edition (2011).

\bibitem{Baumann:2010aa}K. Baumann, C. Guerlin, F. Brennecke and T. Esslinger, Nature \textbf{464}, 1301 (2010).
\bibitem{PhysRevLett.118.073001}
R.~Puebla, M.-J. Hwang, J.~Casanova, and M.~B. Plenio,  Phys. Rev. Lett. \textbf{118} 073001  (2017).

\bibitem{br}D. Braak, Phys. Rev. Lett. \textbf{107} 100401 (2011).
\bibitem{chen}
Q.-H. Chen, C.~Wang, S.~He, T.~Liu, and K.-L. Wang, Phys. Rev. A \textbf{86},  023822 (2012).
\bibitem{jorge}J. Casanova, G. Romero, I. Lizuain, J. J. Garc\'{\i}a-Ripoll, and E.
Solano, Phys. Rev. Lett. \textbf{105}, 263603 (2010).
\bibitem{ion}J. S. Pedernales, I. Lizuain, S. Felicetti, G. Romero, L. Lamata and E. Solano, Sci. Rep. \textbf{5} 15472 (2015).
\bibitem{jin}J.-F. Huang, J.-Q. Liao, L. T., and L.-M. Kuang, Phys. Rev. A \textbf{96}, 043849 (2017).

\bibitem{pj0}J. Peng, Z. Z. Ren, G. J. Guo, G. X. Ju
and X. Y. Guo, Eur. Phys. J. D \textbf{67}, 162 (2013).
\bibitem{guxin}X. Gu, A. F. Kockum, A. Miranowicz, Y. Liu, F. Nori, Phys. Rep. \textbf{718-719}, 1 (2017).
\bibitem{jorge1}J. Casanova, R. Puebla, H. Moya-Cessa and M. B. Plenio, npj Quantum Inf. \textbf{4}, 47 (2018).
\bibitem{KMLSN19} A. F. Kockum, A. Miranowicz, S. De Liberato, S. Savasta, and F. Nori, Nat. Rev. Phys. \textbf{1},19 (2019).
\bibitem{sl}See Suplemental Material for additional details.

\bibitem{ying}Z.-J. Ying, L. Cong, X.-M. Sun, {\it preprint at}	arXiv:1804.08128 (2018).
 \bibitem{chen1}Y.-F. Xie, L. W. Duan, and Q.-H. Chen, Phys. Rev. A \textbf{99}, 013809 (2019).
 \bibitem{pj1}J. Peng \textit{et.al}, J. Phys. A: Math. Theor. \textbf{50} 174003 (2017).
\bibitem{PhysRevA.92.033817}
S.~Felicetti, J.~S. Pedernales, I.~L. Egusquiza, G.~Romero, L.~Lamata,
D.~Braak, and E.~Solano,
Phys. Rev. A \textbf{92}, 033817 (2015).
  \bibitem{pj2}
J.~Peng, \textit{et. al.},
 J. Phys. A: Mathe. and Theor.
 \textbf{47}, 265303 (2014).






\end{thebibliography}

\begin{thebibliography}{11}


\bibitem{PhysRevA.8.25170}
  K.~Hepp and E.~H. Lieb,  Phys. Rev. A  \textbf{8}, 2517 (1973).
  \bibitem{Davis:1957wj}
C.~Davis.
Proc. Amer. Math. Soc. \textbf{8}, 42--44 (1957).
\bibitem{lieb}E. H. Lieb, Commun. Math. Phys. 31, 327 (1973).
\end{thebibliography}
\end{document}